 \documentclass[a4paper,fleqn,usenatbib]{mnras}
\usepackage{newtxtext,newtxmath}

\usepackage[T1]{fontenc}
\usepackage{ae,aecompl}

\usepackage{graphicx}	
\usepackage{amsmath}	
\usepackage{amssymb}	

\newcommand{\sechp}{$\rm sech^2$}
\newcommand{\paperI}{Paper I}
\newcommand{\paperII}{Paper II}
\newcommand{\xu}{X15}

\newcommand{\Oa}{$O8.0-0.1$}
\newcommand{\Ob}{$O9.0+1.0$}
\newcommand{\Oba}{$O8.5+4.0$}
\newcommand{\Oc}{$O14-2.0$}
\newcommand{\Od}{$O10+0.3$}
\newcommand{\Oe}{$O12+4.0$}
\newcommand{\Da}{$D8.5+0.2$}

\newcommand{\Dc}{$D10-0.5$}
\newcommand{\Dd}{$D14+1.5$}

\title[Mapping the Milky Way with LAMOST III]{Mapping the Milky Way with LAMOST III: Complicated spatial structure in the outer disc}

\author[H-F. Wang et al.]{
Hai-Feng Wang,$^{1,2}$\thanks{E-mail: hfwang@bao.ac.cn}
Chao Liu,$^{1,2}$\thanks{E-mail: liuchao@nao.cas.cn}
Yan Xu,$^{1}$
Jun-Chen Wan,$^{1}$
and Licai Deng$^{1}$
\\
$^{1}$ Key laboratory of Optical Astronomy, National Astronomical Observatories, Chinese Academy of Sciences, Beijing, 100012, China\\
$^{2}$ University of Chinese Academy of Sciences, Beijing, 100049, China\\
}

\date{Accepted 2018 April 23. Received 2018 April 13; in original form 2018 March 15}

\pubyear{2017}

\begin{document}
\label{firstpage}
\pagerange{\pageref{firstpage}--\pageref{lastpage}}
\maketitle

\begin{abstract}
We present {complexity} of the Galactic outer disc {by fitting the stellar volume densities of the red giant branch stars with a two-disc component model}. {The discs are confirmed to} extend to $R\sim19$\,kpc. The radial density profile of the discs shows {two breaks at $R\sim11$ and $\sim14$\,kpc, respectively, which} separate the radial profile into three segments with different scale lengths of $2.12\pm0.26$, $1.18\pm0.08$, and $2.72$\,kpc at $R<11$, $11\leq R\leq14$, and $R>14$\,kpc, respectively. The first {break} is likely due to the sudden drop in the radial profile of the thin disc, which may be an evidence of the radial migration. {Beyond $14$\,kpc, the thick disc becomes prominent and the transition from thin to thick disc leads to the second break.} This implies that the geometrically defined thick disc is more {radially} extended than the thin disc. This is also supported by the larger scale length of the thick disc than that of the thin disc. Meanwhile, {the scale height of the thicker component increases from $0.637_{-0.036}^{+0.056}$ at $R=8$ to $1.284_{-0.079}^{+0.086}$\,kpc at $R=19$\,kpc, showing an intensive flared disc}.  Moreover, rich substructures are displayed in the residuals of the stellar density. Among them, the substructures $D14+2.0$ and $O14-1.5$ show a north-south asymmetry, which can be essentially explained by southward shifting of the thick disc. However, no significant overdensity is found for the Monoceros ring. Finally, the thick disc shows a ripple-like feature with unclear origin at $9<R<10.5$\,kpc.
\end{abstract}


\section{Introduction}\label{sec:intro}
The Galactic disc is usually thought to have exponentially declining profiles in both vertical and radial directions~\citep[][etc.]{gilmore1983,juric2008,bovy2012c,blandhawthorn2016,liu2017a}. In the vertical direction, an alternative \sechp\ form is also freguently used as the density profile model~\citep{vanderkruit1988,vanderkruit2011}. 

The Galactic disc can be separated into a thin and a thick components based on the star count derived in the solar neighbourhood \citep{gilmore1983}. \citet{juric2008} found that the thin disc has a smaller scale length than the thick disc, while \citet{bovy2012c} showed opposite trend, i.e. the scale length of the thin disc is larger than that of the thick disc. Moreover, based on the mono-abundance populations, \citet{bovy2012c} argued that the scale length is a function of chemical abundance. If the $\alpha$ abundance is roughly treated as the proxy of age, then the authors showed that the younger population has a larger scale length while the older populations has smaller. Recently, \citet{wan2017} confirmed that the younger red clump population is indeed has a larger scale length than the older red clump population. 

{Although there are variations in the estimated sizes of the thin and thick disc scale heights, most studies agree} with each other that the scale height of the thin disc is around 220--450\,pc, while the value for the thick disc is between 700 and 1200\,pc in the solar vicinity~\citep{blandhawthorn2016}. The scale heights for the younger populations become smaller, implying that the younger populations are thinner than the older ones \citep[][etc.]{liu2012,bovy2012c}. This picture can be explained by the secular heating of the stellar disc due to the scattering of spiral structures and giant molecular clouds~\citep{quillen2001,yu2017}.

Most of the studies of the star count of the Galactic discs are confined within a few kiloparsec around the Sun. \citet[][hereafter \paperI]{liu2017a} extended the star count to the outskirt of the Galaxy with galactocentric distance of about {40}\,kpc. They found that the disc still contributes about 10\% to the surface density profile at about $R=19$\,kpc. This evidence does not agree that the disc truncates at around $14$--$15$\,kpc~\citep{reyle2009,minniti2011}, but is consistent with \citet{carraro2010,carraro2017}, \citet{carraro2015}, and \citet{feast2014} that the disc may extend to larger radius than $\sim20$\,kpc. 

Fitting with a radial exponential profile, \paperI\ obtained that the scale length in the outer disc is only about 1.6\,kpc, much smaller than the values measured in the solar neighbourhood~\citep[see the reviews by][]{rix2013,blandhawthorn2016}. \cite*{liu2017c} later noted that if they confine the surface density profile to the range of $8.5<R<11.5$\,kpc, the scale length becomes $2.37\pm0.02$\,kpc, similar to other works. This implies that the radial density profile of the outer disc may not follow a perfect exponential form, but show a down-bending break at around $R\sim11$\,kpc. It is known that the shape of the radial surface brightness profile of the external galaxies can be roughly classified as three types~\citep{pohlen2006}. It is quite curious which type that the Milky Way belongs to. In principle, the shape of the radial profile may be used to probe the secular evolution of the Galaxy~\citep{debattista2006,roskar2008}.

However, because \paperI\ used the surface density profile in which the vertical details have been marginalised, it may not be the best way to unveil the 2-dimensional (radial and vertical) features in the Galactic outer disc.
Indeed, the radial structure in the outer disc has been found quite complicated such that it is not a simplistic exponential nor a \sechp\ shape, but distorted by flaring, meaning that the scale height increases with $R$. \citet{lopez2002} suggested a new disc model taking into account an exponentially increasing flare along the Galactocentric radius. However, the thick disc is not taken into account in their flare model. Later, \citet{lopez2014} changed the form to quadratic polynomial and allowed for the thick component in their new flare model. They derived a flare for both thin and thick discs with moderate increasing rate in the outskirt of the Galaxy. Using red clump stars, \citet{wan2017} measured the flare of the thin disc without presuming any analytic form. They found that the increasing trend of the flare from $R=9$ to $14$\,kpc is quite similar to \citet{lopez2014}. 

When the number of the observed stars are large and the accuracy of the star counting is improved, the subtle substructures in the vertical stellar density profile can be revealed. \citet{widrow2012} found wave-like oscillations in the residual of the vertical stellar density profile after subtracting a \sechp\ model~\citep[also see][]{johnston2017}. Simulations show that a merging dwarf galaxy can induce perturbation to the stellar disc and can produce similar oscillations in stellar density~\citep{gomez2013,donghia2016,gomez2017}.

The wave-like feature is not only discovered in the solar neighbourhood, but also found in the Galactic outer disc. \citet[][hereafter \xu]{xu2015} found that the star count located at about 2--3\,kpc beyond the location of the Sun is larger in the north of the disc mid-plane than in the south (\emph{north near structure}). The authors also reveal another asymmetric feature at the distance of about 5\,kpc at which the stars become more in the south than in the north (\emph{south middle structure}). Moreover, they found that, at distance of about 8--10\,kpc, the Monoceros ring makes the north more denser than the south again. All these substructures form a wobbly disc in a large range of radii.

The origin of the Monoceros ring reported in various works~\citep{newberg2002, Ive08} is still in debate~\citep{conn2012}. Some works tend to attribute it to a part of the accretion debris from a disrupting satellite~\citep{yanny2003,martin2004}. Some others argued that it may be part of the flare or warp in the outer disc~\citep{momany2006,Lop11}. Recently, \cite{purcell2011} and \cite{laporte2018} found that the perturbation induced by the satellites surrounding the Milky Way can also produce similar substructure. 

\paperI\ found that the Monoceros ring is not statistically significant in the stellar density map covering a large continuous sky area from Galactic mid-plane to high Galactic latitude. It seems that the Monoceros ring is part of the disc. Indeed, the kinematics of the Monoceros ring seems very like the outer disc~\citep{li2012,deboer2018}. The metallicity of this feature is between -0.8 and -1~\citep{juric2008,li2012,conn2012}. Although it is lower than the typical value of the thin disc in the solar neighbourhood, it is similar to the thick disc population {or, after taking into account the abundance gradient, it is also consistent with the regular thin disc}. Note that \citet{yanny2003} estimated the ``preliminary'' metallicity of -1.6 for the Monoceros ring with large dispersion.

In this work, we follow the \paperI\ to map the disc structure in $R$--$Z$ plane from the location of the Sun to about 20\,kpc away from the Galactic centre using the LAMOST red giant branch (RGB) stars. Unlike using the surface density profile as in \paperI, we turn to use the volume density so that the spatial structure of the outer disc in both vertical and radial directions can be probed. Firstly, we measure the vertical structural parameters at different radius slices. Then we investigate how the stellar density profiles and scale heights of different disc components change with radii. 

The paper is organised as below. In section~\ref{sec:data}, we describe the selection of the RGB tracers and briefly introduce how the stellar density is derived. In section~\ref{sec:model}, we build the vertical star count model used in various $R$ slices. In section~\ref{sec:results}, we display the results about the structural features in the outer disc. Then we raise discussions in section~\ref{sec:disc} and briefly summarise in the last section.

\section{Data}\label{sec:data}
\subsection{Sample selection}\label{sec:datasel}
Paper I has shown that the RGB stars are good tracers for the discs as well as the stellar halo. Therefore, we use the RGB stars observed in the LAMOST survey~\citep{cui2012,zhao2012,deng2012} to probe the structures of the outer disc. We select the RGB stars from the LAMOST DR3 catalogue, which contains about 3.2 million stellar spectra with stellar parameter estimates based on the pipeline described by~\citet{luo2015}.   

We start with the selection of the K giant stars following the criteria suggested by~\citet{liu2014}. About 700 thousands K giant stars are selected from the LAMOST DR3. Then the identified red clump stars~\citep{wan2015,tian2017} are removed to obtain the pure RGB stars from the K giant samples. About 570\,000 RGB stars are left after {excluding possible red clump stars}.

The probability density functions of absolute magnitude for the RGB stars are determined using the method developed by~\citet{carlin2015}. Then the distances to the stars are determined from the derived absolute magnitudes combined with the interstellar extinctions estimated from Rayleigh-Jeans colour excess~\citep{majewski2011,zasowski2013}. \citet{wang2017} investigated the external error of the distance by comparing with TGAS data~\citep{gaia2016,astraatmadja2016} and found that the systematic bias is about $-10$\% with random uncertainty of about 20\%.

\begin{figure}
	\centering
 	\includegraphics[scale=0.5]{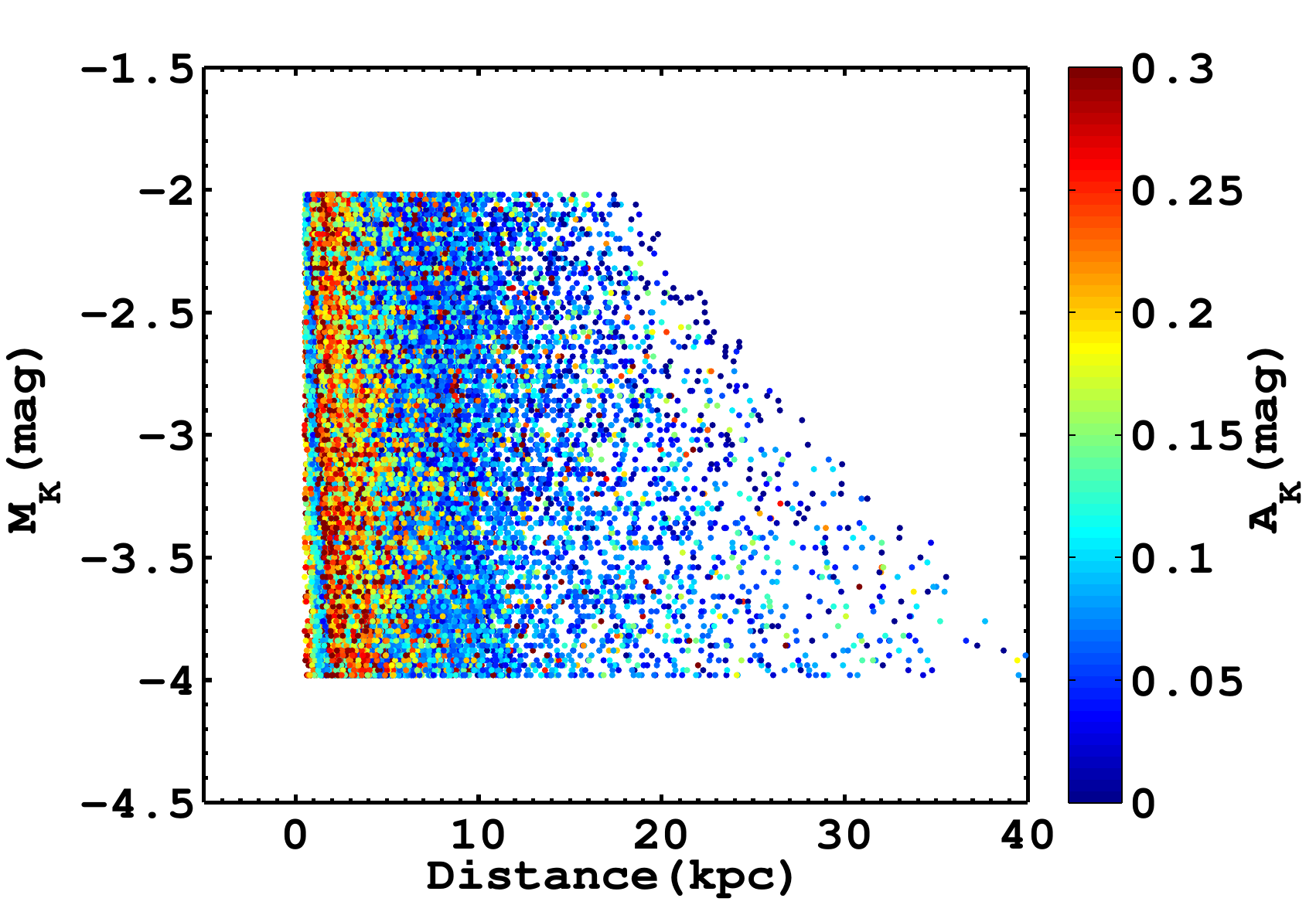}
    \caption{Distribution of the RGB samples in $M_{K}$ vs. distance plane. The colours code the interstellar extinction $A_K$.}
\label{fig:DMK}
\end{figure}

\begin{figure*}
	\centering
	\hspace{1.5cm}
	\includegraphics[scale=0.5]{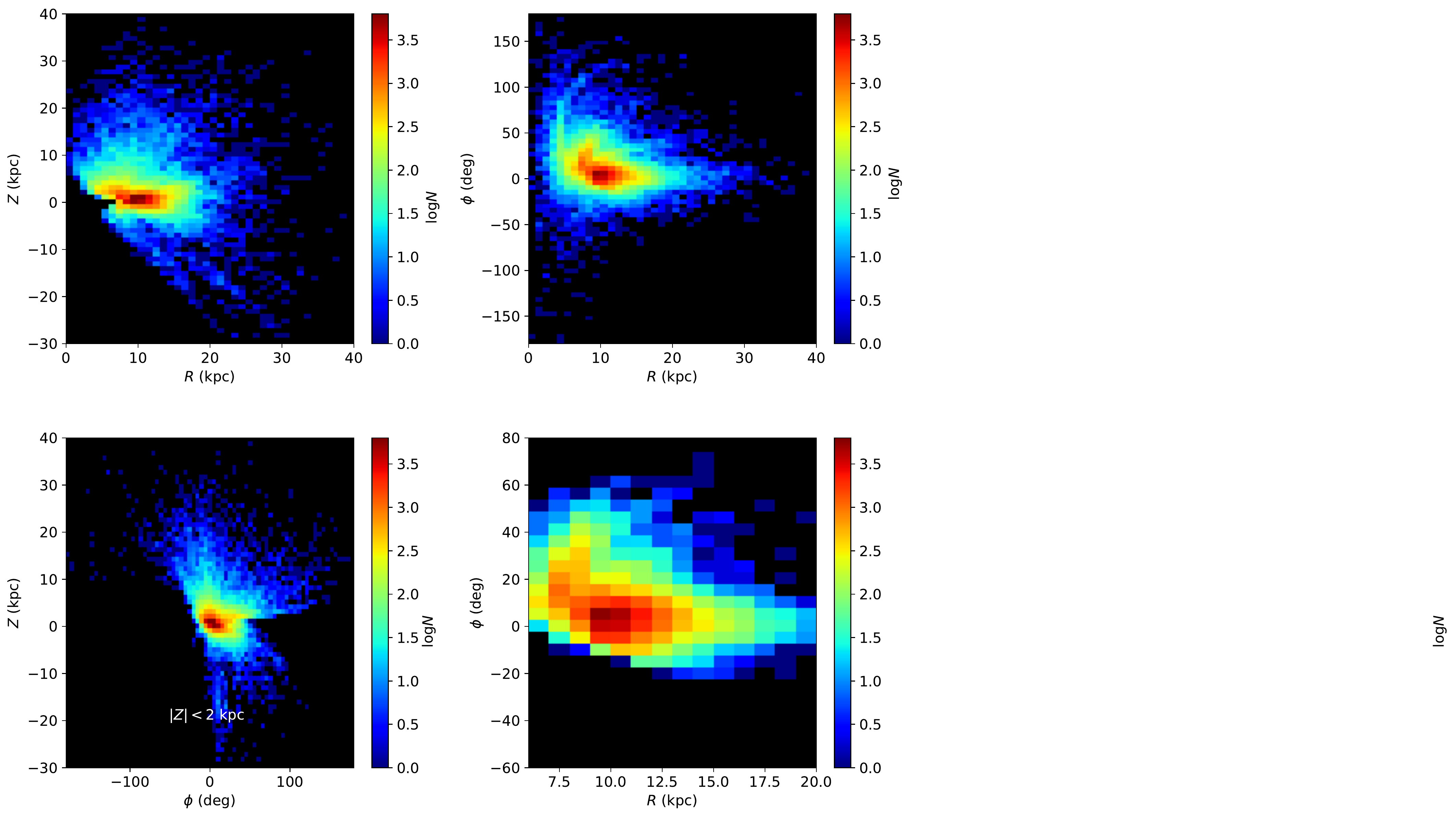}
	\caption{The top-left panel shows the spatial distribution of the RGB stars in $R$--$Z$ plane. The colours code the observed number of stars in logarithmic scale ($\log N$) for each bin. The black regions are not observed  by LAMOST survey. The top-right and bottom-left panels show similar distribution, but in $R$--$\phi$ and $\phi$--$Z$ planes, respectively. $\phi$ is the azimuth angle with the increasing direction same as the direction of the rotation of the Galaxy. The bottom-right panel shows the spatial distribution for the sub-sample of the RGB stars with $|Z|<2$\,kpc in $R$--$\phi$ plane.}\label{fig:3ddist}
\end{figure*}

Although using all RGB stars can extremely increase the number of samples, the different volume completeness for different absolute magnitudes would induce Malmquist bias. Therefore, we select the RGB stars with $-4<M_{K}<-2$\,mag, where $M_{K}$ is the absolute magnitude in 2MASS $K_s$-band. The stars in this range are volume complete within the distance of $\sim20$\,kpc (see Figure~\ref{fig:DMK}). Meanwhile, to ensure that all the stars have reliable 2MASS photometry, we select the RGB stars with $K<14.3$\,mag, which is the limiting magnitude of 2MASS data~\citep{skrutskie2006}. Note that most of the LAMOST RGB stars are within this range, such a cut would not induce substantial selection effect in the samples (see \paperI). Finally, 69\,923 RGB stars are selected for this work.

{Figure~\ref{fig:3ddist} shows the three dimensional distribution of the selected RGB samples in Galactocentric cylindrical coordinates. The direction of the azimuth angle, $\phi$, is chosen to be same as the rotation direction of the Galaxy. Because LAMOST survey only observe the northern sky, the RGB stars are more located in area with positive $Z$.}

{The bottom-right panel shows the spatial distribution in $R$--$\phi$ plane for the sub-sample with $|Z|<2$\,kpc. Although a few stars spread in a wide range of $\phi$, more than 80\% of the RGB samples are concentrated within $\sim20$$^\circ$ in $\phi$ when $R>10$\,kpc. In other word, the RGB samples are mostly concentrated in the Galactic anti-centre (GAC) direction.}

 \subsection{The stellar density map in the outer disc}\label{sec:density}
According to \paperI, the selection effect induced by the target selections and observations can be corrected during the determination of the stellar density. We briefly describe the approach in this section. More details and performance assessment can be referred to \paperI.

Firstly, we assume that the 2MASS photometry is the complete dataset for the stars brighter than $K=14.3$\,mag. Then we assume that, at a given line-of-sight represented by the Galactic coordinates ($l$, $b$) with solid angle $\Omega$, the probability finding a star at distance $D$ in the LAMOST data should be roughly same as in the 2MASS data with similar colour and magnitude, i.e.,

\begin{equation}\label{eq:pp}
p_{ph}(D|J-K,K,l,b)=p_{sp}(D|J-K,K,l,b),
\end{equation}
where $J-K$ and $K$ stand for the colour index and $K_s$-band magnitude in 2MASS, respectively.

Then the stellar densities for photometric data, $\nu_{ph}$, and for spectroscopic data, $\nu_{sp}$, are associated with each other through

\begin{eqnarray}\label{eq:nuPDF}
&\nu_{sp}(D|J-K,K,l,b)=\nonumber\\
&\nu_{ph}(D|J-K,K,l,b)S(J-K,K,l,b),
\end{eqnarray}
where $S$ represents for the selection function of the spectroscopic data.
The stellar density at distance $D$ for the photometric data, which is supposed to be the real density, can be obtained by integrating over colour index and magnitude:
\begin{eqnarray}\label{eq:nuall}
&\nu_{ph}(D|l,b)=\iint\nu_{sp}(D|J-K,K,l,b)\nonumber\\
&S^{-1}(J-K,K,l,b)d(J-K)dK.
\end{eqnarray}

The selection function can be determined by
\begin{equation}\label{eq:S2}
S(J-K,K,l,b)={n_{sp}(J-K,K,l,b)\over{n_{ph}(J-K,K,l,b)}},
\end{equation}
where $n_{sp}$ and $n_{ph}$ are the star counts of the spectroscopic and photometric data in $J-K$ vs. $K$ plane, respectively.

Because the number of spectroscopically observed stars are usually very limited in a line-of-sight and the uncertainty of distance to the stars cannot be ignored, a kernel density estimation (KDE) is applied to derive $\nu_{sp}$ along a line-of-sight, i.e.

\begin{equation}\label{eq:KDE2}
\nu_{sp}(D|J-K,K,l,b)={1\over{\Omega D^2}}\sum_{i}^{n_{sp}(J-K,K,l,b)}{p_i(D)},
\end{equation}
where $p_i(D)$ is the probability density function of $D$ for the $i$th star.
In principle, the stellar density can be derived at any distance along the given line-of-sight, since it is a continuous function of distance based on Eq.~(\ref{eq:KDE2}). However, at the distance in which no sample star is located, the density estimates may not be reliable, according to the tests by \paperI. Thus, we only use the stellar density value at the location of the sample star.

In LAMOST survey, each observation positioning a line-of-sight is denoted as a ``plate'', which has a solid angle of $20$ square degrees. Because of the 4000 fibres installed on the focal plane, each plate can typically observe 2000-3000 stars (most of the rest fibres are for sky background, while a few of them are spire or dead fibres). This allows sufficient statistics to measure one $\nu_{ph}(D)$ for a plate. 
\paperI\ shows that the error of a $\nu_{ph}(D)$ estimate is around $\sigma_\nu/\nu=0.25$. However, because LAMOST survey mostly covers a line-of-sight with multiple plates containing different targets, the stellar density of the line-of-sight is actually measured multiple times with different sample stars. Therefore, the error of the stellar density can be effectively suppressed by averaging over the multiple measurements. 



\begin{figure}
	\centering
 	\includegraphics[scale=0.3]{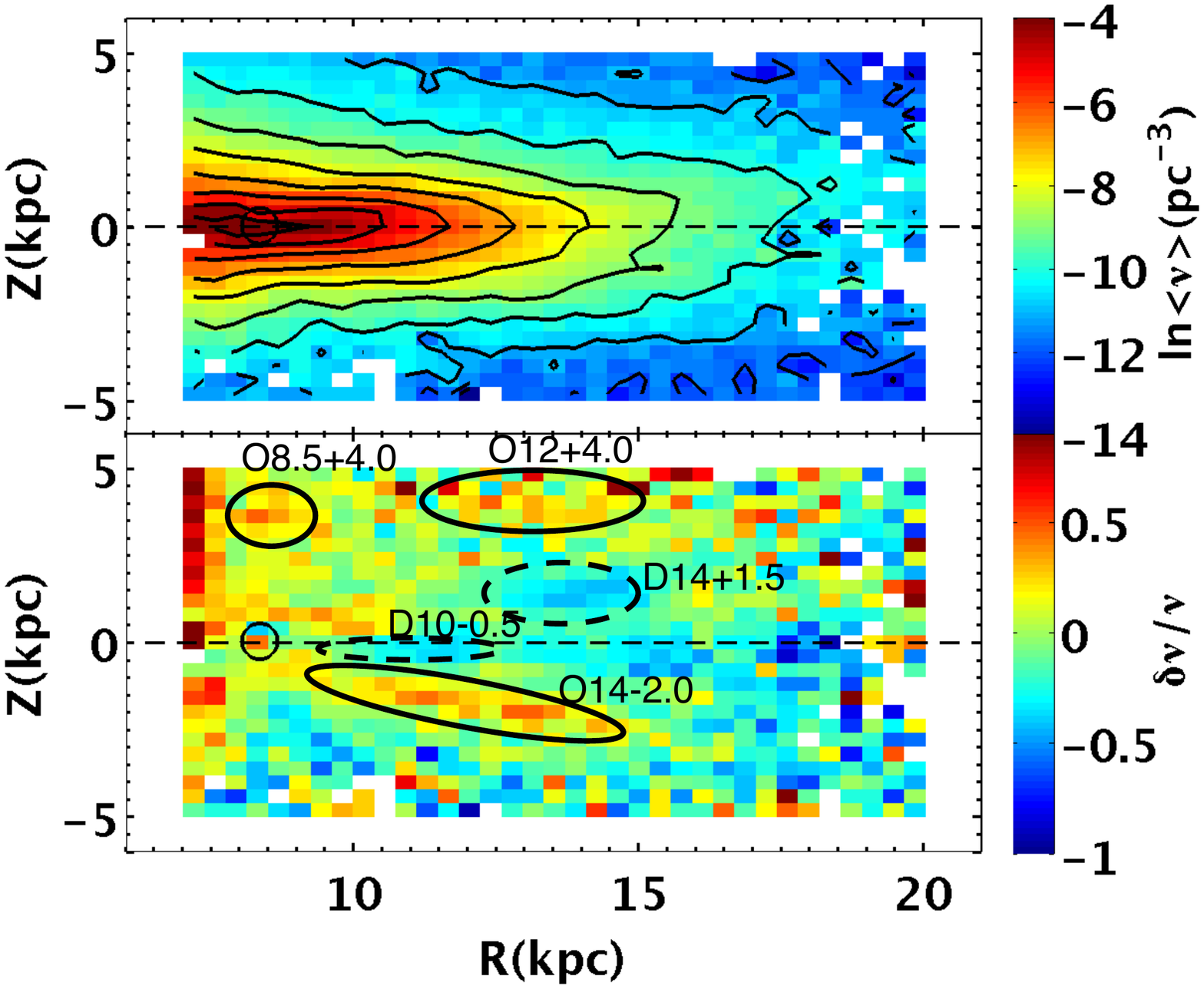}
 	\vspace{0.1cm}
 	\caption{The top panel shows the map of the mean stellar density in $R$--$Z$ plane. The colours code the logarithmic mean stellar density in $\ln({\rm pc}^{-3})$. The small circle indicates the location of the Sun. The bottom panel shows the relative residual map in which colours code the value of $\delta\nu/\nu\equiv(\langle\nu\rangle-\nu_{model})/\nu_{model}$. The black solid and dashed ellipses emphasise the substructures described in section~\ref{sect:substructure}.}
    \label{fig:RZdensityresidual}
\end{figure}

The top panel of Figure~\ref{fig:RZdensityresidual} shows the map of the mean stellar density in $R$--$Z$ plane, where $R$ and $Z$ are galactocentric cylindrical coordinates. {The stellar density, $\nu$, is defined as the number of the RGB stars per pc$^{3}$.} We adopt that the Sun is located at $R_0=8.34$\,kpc \citep{reid2014} and $Z_0=0.027$\,kpc~\citep{chen2001}. As in \paperI, the RGB stars selected for this work display similarly large disc, which does not truncate within $R\sim20$\,kpc. 

{\paperI\ has compared the derived interstellar extinction with the 3-D extinction model provided by \citet{green2014,green2015} at a few lines-of-sight with high extinction in the GAC region. It shows that the two extinctions are consistent with each other within an uncertainty of $A_K\sim0.2$\,mag, which can induce less than 10\% uncertainty in distance. Therefore, we infer that the uncertainty of extinction may not substantially affect the stellar density profiles. However, for a few distant and faint (i.e. $K\sim14$\,mag) RGB stars, high extinction may lead to slight volume incompleteness. Those faint stars with larger extinction may have apparent magnitude fainter than $K=14.3$\,mag and hence are cut off. Because high interstellar extinction areas are mostly located at southern Galactic disc due to the nearby star forming regions e.g. Orion, Taurus etc., this may lead to some north-south asymmetry at large $R$. Indeed, the top panel of Figure~\ref{fig:RZdensityresidual} does show a spike-like feature at around $Z\sim-1$\,kpc and $R>15$\,kpc, which is likely the effect of the extinction.}

In next sections, we quantify the stellar density profile by modelling the vertical density profile at various radius bins. The relative residual map shown in the bottom panel of Figure~\ref{fig:RZdensityresidual} will be discussed in section~\ref{sect:substructure}.

\section{Star count model}\label{sec:model}
At a given $R$ bin, the vertical stellar density profile  is composed of three components, the thin and thick disc and the stellar halo. We adopt the \sechp\ model~\citep{vanderkruit1988} for the thin and thick discs and a power-law for the halo. 

For the thin disc component, the vertical density profile at given $R$ can be written as
\begin{equation}\label{eq:thin_nu}
\nu_{\rm thin}(Z|R)=\nu_0(R)(1-f_t(R)-f_h(R)){\rm sech}^2\left({Z\over{2h_{z1}(R)}}\right),	
\end{equation}
where $\nu_0(R)$ is the total stellar density at $Z=0$, $f_t(R)$ and $f_h(R)$ are the fractions of the thick disc and halo component at $Z=0$, respectively. And $h_{z1}(R)$ is the scale height of the thin disc. All these parameters are functions of $R$ without any pre-defined analytical form.

For the thick disc, it becomes
\begin{equation}\label{eq:thick_nu}
\nu_{\rm thick}(Z|R)=\nu_0(R)f_t(R){\rm sech}^2\left({Z\over{2h_{z2}(R)}}\right).
\end{equation}
Similar to the thin disc, the scale height of the thick disc, $h_{z2}(R)$, is also a function of $R$ without pre-defined form.

{Note that the stellar warp are neglected in the thin and thick disc models. As shown in the bottom-right panel of Figure~\ref{fig:3ddist}, the RGB stars are mostly concentrated along GAC region. According to \citet{lopez2002}, the GAC direction is nearly overlapped with the line-of-node of the stellar warp, meaning that the stars located around GAC is not sensitive to the warp. We further quantify the effect of the warp at $R\sim14$\,kpc as an instance. Adopting the warp model from \citet{lopez2002}, we find that the warp induced vertical offsets is around $0.01$\,kpc for the selected RGB stars located at around $R=14$\,kpc. Compared to the scale height at the same radius, which is $0.36$\,kpc for the thin disc (see Table~\ref{tab:bestfitparams}), neglecting the warp in the disc models may not substantially change the result.}

For the halo, the density model can be written as
\begin{equation}\label{eq:halo_nu}
  \nu_{\rm halo}(Z|R,n,q) = \nu_0(R)f_h(R)\left({R \over \sqrt{R^2 +
(Z/q)^2}}\right)^{n},
\end{equation}
where $n$ and $q$ are the power index and axis ratio of the stellar halo, respectively, providing that the halo is axisymmetric. Since we focus on the shape of the disc in this work, we fix the parameters of the halo, $n$ and $q$, for simplification. Hence, $\nu_0$, $f_t$, $h_{z1}$, $h_{z2}$, and $f_h$ are the free parameters in the model. 

Combining Eqs~(\ref{eq:thin_nu})-(\ref{eq:halo_nu}), the total stellar density model becomes
\begin{eqnarray}\label{eq:nu_tot}
&\nu_{\rm model}(Z|R,\nu_0,h_{z1},h_{z2},f_t,f_h)=\nonumber\\
&\nu_{\rm thin}(Z|R)+\nu_{\rm thick}(Z|R)+\nu_{\rm halo}(Z|R,n,q).	
\end{eqnarray}

\citet[][hereafter \paperII]{xu2017} showed that $q$ is correlated with the galactocentric radius. Within $R<20$\,kpc, $q$ changes from $\sim0.5$ to 0.8. \paperII\ also gives the best fit power index of $5$ with variable $q$. We adopt the result from \paperII\ and fix $n$ at $5$. We adopt a variable $q$ following the empirical relationship with galactocentric radius $r$ suggested by \paperII. Hence $q$ becomes
\begin{equation}\label{eq:halo_nu2}
q=q(r)=q(\sqrt{R^2+Z^2}).
\end{equation}

To derive all the free parameters mentioned in Eq~(\ref{eq:nu_tot}) at each $R$ bin, we firstly set up the histogram of the mean vertical stellar density along $Z$ grid. Then we obtain the likelihood distribution of the vertical stellar density profile as
\begin{eqnarray}\label{lnlike}
  &\mathcal{L}(\{\nu_{\rm obs}(Z_i|R)\}|\nu_0,h_{z1},h_{z2},f_t,f_h) =\nonumber\\ &\prod_i\exp\left[-{1\over{2}}(\nu_{\rm obs}(Z_i|R)-\right.\nonumber\\
  &\left.\nu_{\rm model}(Z_i|R,\nu_0,h_{z1},h_{z2},f_t,f_h)^2\right],
\end{eqnarray}
where $Z_i$ is the $i$th point of the $Z$ grid. It is then sampled 
with a Markov chain Monte Carlo (MCMC) simulation provided by \textsc{emcee}~\citep{foreman2013}. The best-fit values of the free parameters are determined by the peaks of the marginalised likelihood distributions. The uncertainties of the estimates are determined using the 15\% and 85\% percentiles of the MCMC samples.

\section{Results}\label{sec:results}
\begin{figure}
	\centering
	\begin{minipage}{9cm}
	\centering
	\includegraphics[scale=0.6]{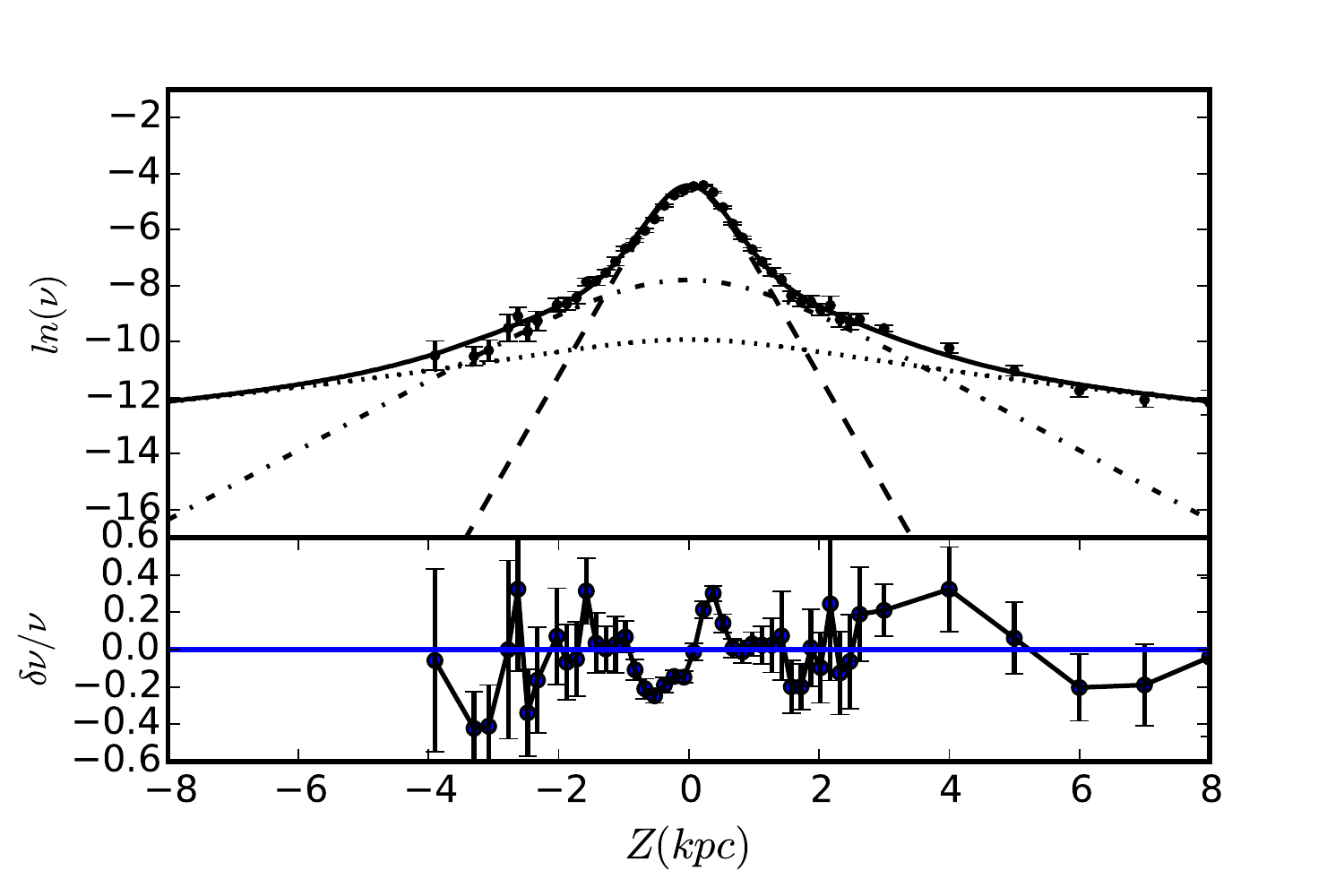}
	\includegraphics[scale=0.27]{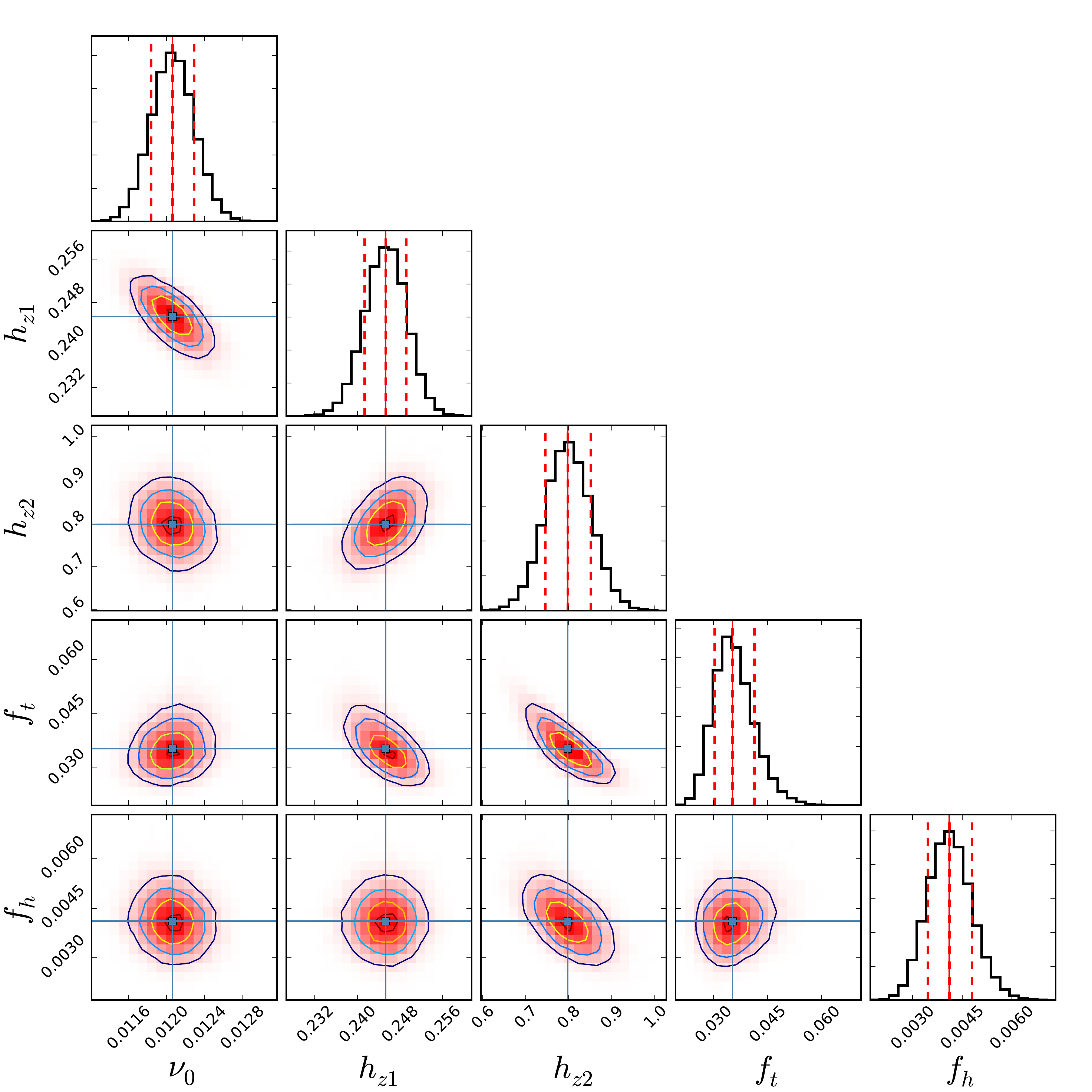}
	\end{minipage}
 	
 	\caption{The main part of the top panel shows observed vertical density profile (black dots with error bars) and the best-fit model (black solid lines) for the bin of $R=10$\,kpc. The black dashed, dash-dotted, and dotted lines indicate the best-fit thin disc, thick disc, and halo, respectively. The relative residuals, $\delta\nu/\nu=(\nu_{obs}-\nu_{model})/\nu_{model}$, is displayed at the bottom of the top panel. The bottom panel shows the likelihood distribution of the parameters ($\nu_0$, $h_{z1}$, $h_{z2}$, $f_t$,$f_h$) drawn from the MCMC simulation.}\label{fig:mcmcR10}
\end{figure}

\begin{figure*}
	\centering
	\includegraphics[scale=0.42]{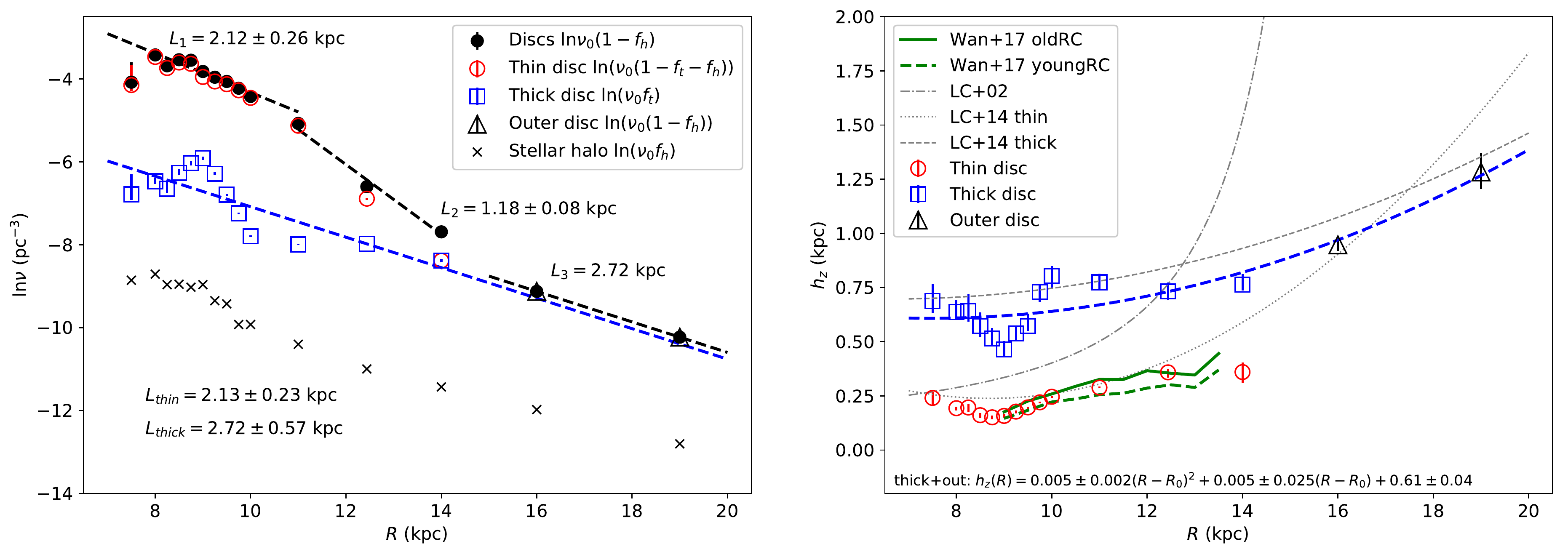}
	\caption{The left panel displays the radial density profiles at $Z=0$\,kpc for different components. The black dots represent for the total radial profile of the disc components, i.e. $\ln(\nu_0(1-f_h))$. The red circles and blue rectangles show the radial profiles of the thin and thick discs, i.e. $\ln(\nu_0(1-f_t-f_h))$ and $\ln(\nu_0f_t)$, respectively, within $R=14$\,kpc. The black triangles display the radial densities of the outer disc, i.e. $\ln(\nu_0(1-f_h))$. As a comparison, the density profile of the halo is displayed as black crosses. The total radial profile of the disc components is fitted with three segments of exponentials and displayed with three black dashed lines. The radial profile of the thick disc is fitted with an exponential and displayed as a blue dashed line.
	The right panel shows the scale heights of the thin, thick and outer discs with the red hollow circles, blue rectangles, and black triangles, respectively. The green solid and dashed lines indicate the scale heights for the old and young red clump stars from~\citet{wan2017}. The grey dash-dotted line indicates the flare model from~\citet{lopez2002}. And the grey dotted and dashed lines show the flare models for the thin and thick discs, respectively, from~\citet{lopez2014}. The blue dashed line shows the quadratic flare model derived from the thick+outer discs. 
	}\label{fig:bestfitparams}
\end{figure*}

We separate the RGB samples into $R$ bins centred at $R=7.5$, $8$, $8.25$, $8.5$, $8.75$, $9$, $9.25$, $9.5$, $9.75$, $10$, $11$, $12.44$, $14$, $16$, and $19$\,kpc. The width of bins is $0.25$ kpc for those with $R\leq10$\,kpc, and enlarged to $1.75$\,kpc at $R=11$\,kpc, $1.125$\,kpc at $R=12.44$, $2$\,kpc at $R=14$ and $16$\,kpc, and $4$\,kpc at $R=19$\,kpc. {The principles about the separation of the bins along $R$ is as below. First, we expect that the bin size is as small as possible to obtain better resolution in $R$. Second, we need to keep as many as possible stars in each bin so that we can have good coverage in $Z$ direction. For most of the bins with $R<10$\,kpc, we tested several solutions and find the bin size of $0.25$ kpc can comply with the statistical requirement. The size is much smaller compared to the scale length. And the numbers of the samples in these bins are larger than 1000. Meanwhile, for the bins at $R>10$\,kpc, we gradually enlarge the bin sizes to include more stars so that there are sufficient stars covering $Z$ direction. In section~\ref{sec:systematics}, we tested that even we set bins with larger size, it would not bring substantially systematical bias in the disc modelling.}

\begin{table*}
\caption{The best-fit parameters of the models of the vertical density profile at different $R$ bins.}\label{tab:bestfitparams}
\centering
\begin{tabular}{c|c|c|c|c|c}
\hline\hline
	$R$ & $\ln\nu_0$ & $h_{z1}$ & $h_{z2}$ & $f_t$ & $f_h$\\
	(kpc) & ($\ln{\rm pc^{-3}}$) & (kpc) & (kpc) &  & \\
	\hline
7.50 & $-4.072_{0.141}^{0.477}$ & $0.241_{0.035}^{0.027}$ & $0.688_{0.055}^{0.077}$ & $0.066_{0.009}^{0.036}$ & $0.008_{0.055}^{0.077}$\\
8.00 & $-3.418_{0.066}^{0.076}$ & $0.193_{0.011}^{0.007}$ & $0.637_{0.036}^{0.056}$ & $0.047_{0.006}^{0.012}$ & $0.005_{0.001}^{0.001}$\\
8.25 & $-3.681_{0.109}^{0.195}$ & $0.197_{0.019}^{0.016}$ & $0.642_{0.049}^{0.077}$ & $0.051_{0.009}^{0.019}$ & $0.005_{0.001}^{0.002}$\\
8.50 & $-3.535_{0.057}^{0.101}$ & $0.161_{0.013}^{0.010}$ & $0.573_{0.052}^{0.063}$ & $0.065_{0.011}^{0.021}$ & $0.004_{0.001}^{0.001}$\\
8.75 & $-3.546_{0.061}^{0.049}$ & $0.151_{0.009}^{0.008}$ & $0.515_{0.036}^{0.049}$ & $0.083_{0.013}^{0.018}$ & $0.004_{0.001}^{0.001}$\\
9.00 & $-3.817_{0.038}^{0.041}$ & $0.158_{0.008}^{0.008}$ & $0.464_{0.028}^{0.039}$ & $0.122_{0.018}^{0.027}$ & $0.006_{0.001}^{0.001}$\\
9.25 & $-3.957_{0.028}^{0.035}$ & $0.178_{0.008}^{0.005}$ & $0.538_{0.035}^{0.034}$ & $0.097_{0.016}^{0.013}$ & $0.005_{0.000}^{0.000}$\\
9.50 & $-4.058_{0.025}^{0.016}$ & $0.197_{0.006}^{0.004}$ & $0.572_{0.022}^{0.058}$ & $0.065_{0.006}^{0.017}$ & $0.005_{0.001}^{0.000}$\\
9.75 & $-4.223_{0.021}^{0.018}$ & $0.221_{0.004}^{0.004}$ & $0.730_{0.047}^{0.036}$ & $0.049_{0.005}^{0.008}$ & $0.003_{0.000}^{0.001}$\\
10.00 & $-4.420_{0.017}^{0.020}$ & $0.246_{0.005}^{0.003}$ & $0.804_{0.060}^{0.045}$ & $0.034_{0.004}^{0.007}$ & $0.004_{0.001}^{0.001}$\\
11.00 & $-5.070_{0.012}^{0.011}$ & $0.289_{0.003}^{0.004}$ & $0.774_{0.038}^{0.040}$ & $0.054_{0.005}^{0.009}$ & $0.005_{0.000}^{0.000}$\\
12.44 & $-6.588_{0.030}^{0.019}$ & $0.359_{0.023}^{0.016}$ & $0.732_{0.041}^{0.041}$ & $0.249_{0.034}^{0.062}$ & $0.012_{0.001}^{0.001}$\\
14.00 & $-7.667_{0.042}^{0.043}$ & $0.360_{0.049}^{0.045}$ & $0.763_{0.026}^{0.048}$ & $0.489_{0.064}^{0.085}$ & $0.023_{0.002}^{0.002}$\\
	\hline
	$R$ & $\ln\nu_0$ & \multicolumn{3}{c}{$h_{z1}$}  & $f_h$\\
	(kpc) & ($\ln{\rm pc^{-3}}$) & \multicolumn{3}{c}{(kpc)}  & \\
	\hline
	16.00 & $-9.070_{0.040}^{0.042}$ & \multicolumn{3}{c}{$0.947_{0.027}^{0.027}$} & $0.055_{0.005}^{0.008}$\\
19.00 & $-10.157_{0.058}^{0.072}$ & \multicolumn{3}{c}{$1.284_{0.079}^{0.086}$} & $0.071_{0.018}^{0.024}$\\
\hline\hline
\end{tabular}	
\end{table*}

The {performance} of the MCMC simulations {is demonstrated in Figure~\ref{fig:mcmcR10}, which displays the simulation at $R=10$\,kpc}. The best-fit parameters for different $R$ bins are listed in Table~\ref{tab:bestfitparams}. 

Beyond $R=14$\,kpc, the vertical stellar densities do not prefer to two disc components such that the MCMC simulation can not converge to meaningful parameter values. Therefore, we only apply one disc component to the model for $R>14$\,kpc. 

At $R>14$\,kpc, the scale heights of the only disc component is comparable to a typical thick disc at $R<14$\,kpc. It can be either the extension of the thick disc population or the outskirt of the substantially flared thin disc population. We simply denote it as the \emph{outer} disc to avoid preconception of which population it belongs to. The nature of the outer disc will be discussed in next section. 

{As discussed in section~\ref{sec:density}, at $R>15$\,kpc, the volume incompleteness induced by the extinction may affect the density profiles at around $Z\sim-1$\,kpc. However, at these large radii, the disc component is quite thick with scale heights larger than $\sim1$\,kpc. Moreover, extinction does not influence the stellar density in the north of the disc. Therefore, extinction would not significantly change the result of the model fitting, although the residual of the model may be enlarged.}

Because that the observed data do not well constrain the stellar density at $R=7.5$\,kpc, we exclude this data point in next sections.

\subsection{The radial density profiles of the discs}\label{sect:radialprofile}

\subsubsection{Type II+III disc profile}\label{sect:drop}
The left panel of Figure~\ref{fig:bestfitparams} shows the resulting radial density profiles at $Z=0$ for different components.

The total radial stellar density for the discs (the black dots) shows two breaks: a down-bending break at around $R=11$\,kpc and an up-bending break located at $R\sim14$\,kpc. As a reference, the radial profile of the stellar halo (black crosses) does not show similar breaks. Therefore, the two breaks only occur in the discs. 

The two breaks separate the radial density profile into three segments: $R=8$--$11$, $11$--$14$, and $16$--$19$\,kpc. They are separately fitted with exponential models and the best-fit scale lengths are $L_{1}=2.12\pm0.26$, $L_{2}=1.18\pm0.08$, and $L_{3}=2.72$\,kpc, respectively.
Because that the last segment contains only two data points, the error of $L_{3}$ is not provided.

The first down-bending break is consistent with \citet{liu2017c}, who found a similar but marginal break at around $11$\,kpc from the surface density profile derived by \paperI. The authors obtained the scale length of the disc as $2.37$\,kpc within $R=8.5$--$11.5$\,kpc, which also agrees with this work.

It is noted that such a double-break radial profile is not unusual in external late-type galaxies. \cite{pohlen2006} shows that NGC 2701, NGC4273, NGC4904, and NGC5147 have a down-bending break followed with a up-bending break, which are very similar to the radial profile shown in the left panel of Figure~\ref{fig:bestfitparams}. The authors classify such kind of profile as type II+III. %

While the radial profile of the thick disc (blue rectangles) is quite flat in the range $11<R<14$\,kpc, the radial density profile of the thin disc (red circles) shows an abrupt drop-down at $R>11$\,kpc and then fades away at larger radii. Therefore, it is likely that the first break at $R\sim11$\,kpc is due to the fading away of the thin disc. 

Radial migration may lead to the first drop-down in the radial profile of the thin disc. \citet{roskar2008} pointed out that the radial density profile of the disc experienced radial migration will exhibit a down-bending break in the outskirt of the disc, which is quite similar to the radial profile of the thin disc. Therefore, the down-bending break shown at $R\sim11$\,kpc may reflect the radial migration occurred in the thin disc.

Although the radial density profile of the thick disc shows a ripple-like feature at $9<R<10$\,kpc (will be discussed in section~\ref{sec:ripple}), it roughly follows an exponential-like profile with larger uncertainty. The best-fit scale length of the thick disc within $R=14$\,kpc is $L_{thick}=2.72\pm0.57$\,kpc. When the exponential profile is extended to $16\leq R\leq19$\,kpc, it well overlaps with the radial densities of the outer disc. Thus, it seems that the outer disc is the extension of the thick disc. While the radial profile of the thin disc gradually fades away at $11<R<14$\,kpc, the thick disc component becomes prominent beyond $14$\,kpc. This transition may lead to the second up-bending break at $R\sim14$\,kpc.

It is worthy to point out that the above analysis is phenomenological. The consistency of the radial profile between the thick and the outer discs does not necessarily mean that they are from the same stellar population. Consequently, other explanations about the two breaks in the radial profile can not be ruled out. One scenario is that the radial migration induced by the minor mergers in the outskirt of the galaxy can significantly affect the stellar distribution in the outer disc, according to the simulations conducted by~\citet{bird2012}. However, these authors did not quantify how the radial density profile change with the perturbation of the minor mergers. Hence, it is not clear whether perturbation from the merging satellites can lead to the up-bending break. 
In the case of external galaxies, \citet{pohlen2006} also suggested that interaction with nearby companions may be responsible for the complicated features in some of the Type II+III galaxies.

\subsubsection{The scale lengths of the thin and thick discs}\label{sect:scalelength}
The scale length of the thin disc within $R\leq11$\,kpc is $L_{thin}=2.13\pm0.23$\,kpc, significantly smaller than that of the thick disc, which is $L_{thick}=2.72\pm0.57$\,kpc. 
{We compare our result with previous works, mainly \citet{juric2008} and \citet{bovy2012c}.}

{\citet{juric2008} used the photometric main-sequence stars with various colour indices observed by SDSS survey and fitted the data with an axisymmetric  star count model containing two disc components. The biggest difference between~\citet{juric2008} and this work is that they assume that the radial density profiles of the two discs follow exponential shapes, while we do not add any assumption on the shape of the radial density profile. This means that, unlike this work, they did not take into account the flare in their model.}

{Since in this work the thin and thick discs essentially show exponential-like radial density profile at $R\leq11$\,kpc, we can compare the scale lengths for the two disc components with these authors at this range of radii. \citet{juric2008} suggested that the scale lengths are 2.6 and 3.6\,kpc for the thin and thick disc, respectively, which are slightly larger than our result. Although these values show systematics to some extent, our results well agree with these authors that the scale length of the thick disc is larger than that of the thin disc.}
The ratio of the scale length of the thick to thin disc is $\sim1.3$, in agreement with the typical ratios found in the external galaxies~\citep{yoachim2006}. The larger scale length of the thick disc also supports that the thick disc dominate the Galactic outskirt and produces the up-bending break when the thin disc fades away.

{\citet{bovy2012c} separated the SDSS observed G-dwarf stars into ``mono abundance sub-populations'', i.e. group the stars with similar [Fe/H] and [$\alpha$/Fe]. Then they fitted the stellar density for each mono abundance population with a single disc component. Both the radial and vertical density profiles of the disc follow exponential. Again, these authors did not take into account the flare of the discs. It is noted that there is neither thin nor thick disc defined by geometry in their model. The high-$\alpha$ abundance populations show larger scale heights and are treated as the thicker component, while the low-$\alpha$ abundance populations display smaller scale heights and are considered as the thinner component ~\citep[for more details about different definitions of the thick disc, please refer to][]{martig2016}. Adopting the definition of the thinner and thicker discs based on the chemical abundance, these authors show that the thicker disc has a smaller scale length and the thinner disc has a larger scale length. Recently, ~\citet{bensby2017}, who also follow the chemical definition of the thin/thick disc, exhibited similar result using high-resolution spectroscopic data.}

{ It is clear that the geometrically defined thick disc, such as \citet{juric2008} and this work, shows larger scale length than the thin disc, while the chemically defined thin/thick discs show opposite trend.}

The discrepancy implies that the geometrically defined thick disc in the larger radii may have lower $\alpha$-abundance and thus it may be treated as the chemically defined thin disc with significant flare. If this is true, then the formation of the flare may be as described as~\citet{minchev2015}.

It is clear that the key to better understand the nature of the double-break in the disc radial profile and the origin of the thin/thick disc in the Galactic outskirt is the stellar chemical abundance, which would be helpful in discriminating the stellar populations. Note that \citet{ho2017} has provided chemical abundances for the LAMOST DR2 data, it is worthwhile to further investigate this issue with chemistry in future works.

\subsection{The flare}\label{sect:flare}
\begin{figure}
\centering 
\includegraphics[scale=0.55]{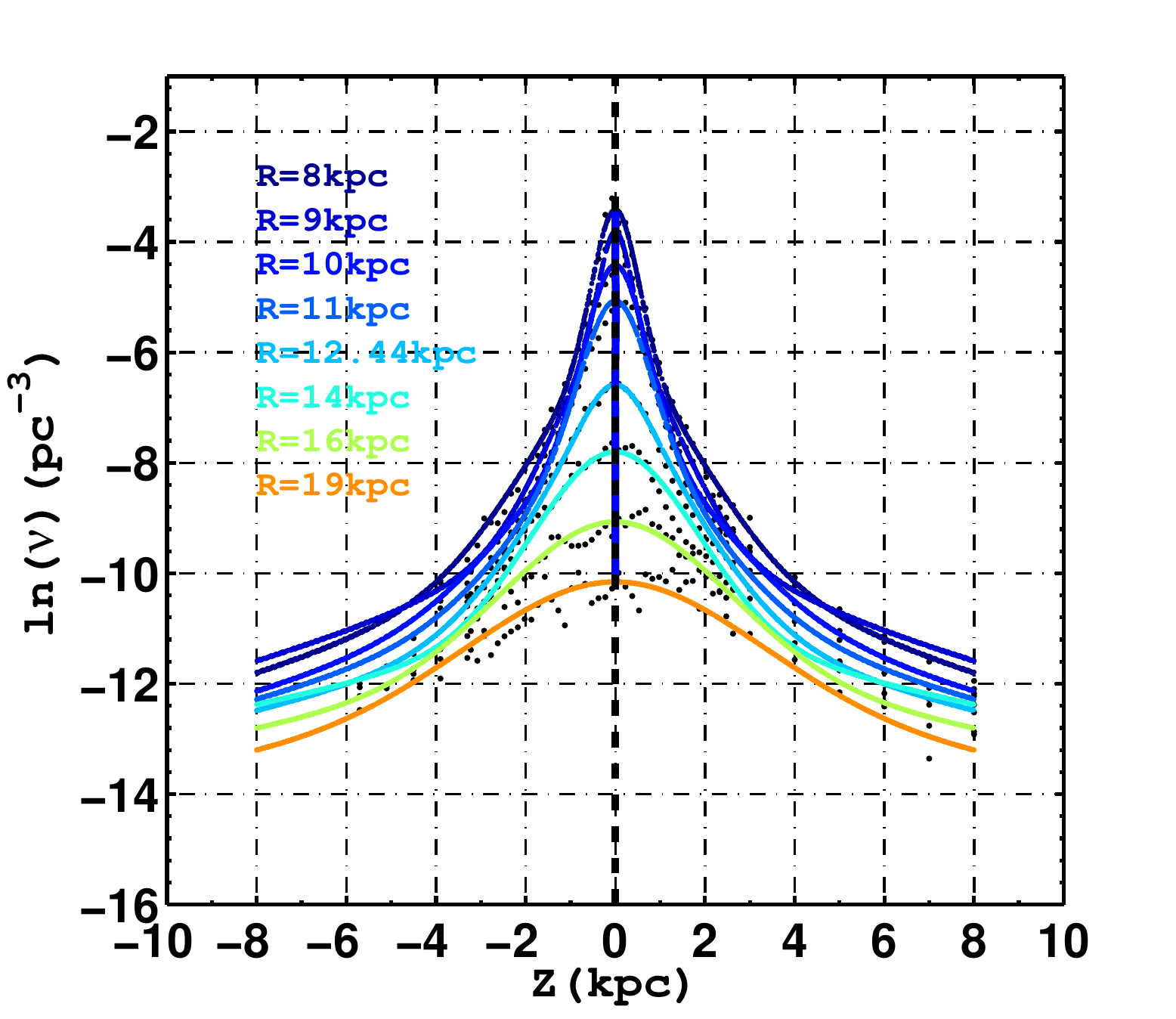}
\caption{The best-fit models of the vertical density profiles at different radii are shown with coloured lines, in which different colours stand for different $R$. The black dots display the observed data points at various $R$ bins.}
\label{fig:flare}
\end{figure}
The flare of the disc can be qualitatively demonstrated by piling the vertical stellar density profiles of different $R$ bins together as shown in Figure~\ref{fig:flare}. It is obviously seen that the vertical density profiles are broadened with increasing $R$, implying that the Galactic disc is substantially thickened with $R$.
However, it is not obvious that whether the flare is mainly contributed by the thin or the thick disc. To better discriminate this, we display the scale heights for the thin, thick, and outer discs with the red hollow circles, blue hollow rectangles, and black hallow triangles, respectively, in the right panel of Figure~\ref{fig:bestfitparams}. 

\subsubsection{The flared thin disc}\label{sec:flarethindisc}
For the thin disc, the scale height slightly declines from $8$ to about $9$\,kpc. Then it mildly increases from $9$ to $\sim12$\,kpc. At $14$\,kpc, the scale height is essentially same as at $12.44$\,kpc, meaning that the thin disc is not further flared at this radius.

{We compare the flare with previous works. \citet{lopez2002} used photometrically selected red clump stars as the tracers of the flare. The authors empirically assumed that the disc is thickened with radius exponentially. Compared to their flare model (which is indicated with a grey dot-dashed line in the right panel of Figure~\ref{fig:bestfitparams}), the scale height trends in this work is much flatter.}

{\citet{lopez2014} turned to use F/G main-sequence stars and adjusted the flare model to quadratic model, which increases quite mildly. Meanwhile, they also take into account the thick disc in their new flare model} (the grey dotted and dashed lines for the thin and thick disc, respectively). Although our result is roughly consistent with ~\citet{lopez2014} within $R<12$\,kpc, the scale height at $R=14$\,kpc is substantially smaller than their quadratic model. 

We compare the flared scale heights for the RGB stars with those for the red clump stars derived by~\citet{wan2017} (the green solid and dashed lines for the old and young populations, respectively). {\citet{wan2017} adopted the same approach as this work to derive the stellar density for the LAMOST red clump stars. We} find that it is quite consistent with the one for the old red clump stars (the green solid line). This is reasonable since the averaged age of the RGB stars is around 3--6\,Gyr~\citep[see Figure 7 in][]{ho2017}, similar to the old red clump stars, which is about 4\,Gyr~\citep{tian2017}.

{\citet{wan2017} shows flatter trend in radial variation of scale height, similar to this work, with essentially same type of tracers as~\citet{lopez2002}. This implies that the exponential flare model may not match the observed data.}

To investigate whether the outer disc is associated with the thin disc in terms of flare, we attempt to combine the scale heights of the two components together. 


The scale height of the outer disc at $16$\,kpc is significantly larger than that of the thin disc at $14$\,kpc by a factor of $2.6$ and hence shows an abrupt jump from $R=14$ to $16$\,kpc. Consequently, the scale height of the thin disc does not smoothly transit to that of the outer disc. This can not be explained by the current theories of the flaring. Theoretically, the flare may be induced by the interaction of the merging satellites~\citep{kazantzidis2008,villalobos2008,bournaud2009} or the secular evolution of the disc~\citep{minchev2012,minchev2015}. Both scenarios expect that the scale height of the flared disc should smoothly increases with radii. Therefore, either the outer disc is not from the same population as the thin disc or there should be another channel to explain the origin of the flare considering the discontinuity.

\subsubsection{The flared thick disc}\label{sec:flarethickdisc}
In the right panel of Figure~\ref{fig:bestfitparams}, the scale height of the thick disc shows an oscillation at around $9$--$10$\,kpc, similar to the stellar density shown in the left panel. We will discuss it in section~\ref{sec:ripple}.
Other than this ripple-like feature, the thick disc does not show substantial flare within $14$\,kpc. However, if we combine the outer with the thick disc, their scale heights can be well fitted with a quadratic model described as below:
\begin{eqnarray}\label{eq:flare_thickout}
h_z(R)&=(0.007\pm0.001)(R-R_0)^2\\\nonumber
&-(0.028\pm0.008(R-R_0)	+(0.70\pm0.01),
\end{eqnarray}
which is displayed as the blue dashed line in the panel. This model is qualitatively in agreement with the thick disk flare model from~\citet{lopez2014} (the grey dotted line) with a slight systematic shift. Although the flare does not necessarily increase as a quadratic polynomial, the observed data well fitted with the continuous model implies that the thickening trend is quite smooth without discontinuity. 

Therefore, it seems that the outer disc population may not belong to the thin disc population, but more likely an extension of the thick disc.
Certainly the conclusion is not exclusive, other explanations can not be ruled out.

\subsection{Substructures}\label{sect:substructure}

\begin{table*}
\caption{List of the substructures.}\label{tab:substructure}
\centering
\begin{tabular}{c|c|c|p{3cm}}
\hline\hline
Name & Location & max. $\delta\nu/\nu$* & Comments\\	
\hline
\Ob\ & $R=8$--$9.5$\,kpc, $Z\sim1$\,kpc &0.28& substructure found by \citet{juric2008}\\
\Od\ & $R=9.5$--$11$\,kpc, $Z=0$--$0.5$\,kpc &0.32& north near structure in \xu\\
\Oc\ & $R=8.5$--$15$\,kpc, $Z=-1$--$-2.5$\,kpc &0.37& south middle structure in \xu\\
\Dd\ & $R=12$--$15$\,kpc, $Z\sim1.5$\,kpc &-0.37& new substructure located in the opposite side of the south middle structure.\\
\Da\ & $R=8$--$9$\,kpc, $Z=0$--$0.5$\,kpc &-0.42& consistent with~\citet{widrow2012}\\

\Oba\ & $R=8$--$9.5$\,kpc, $Z\sim4$\,kpc &0.50& Virgo overdensity?\\
\Oa\ & $R=8$--$8.5$\,kpc, $Z=-0.1$--$0$\,kpc &0.90& effect of incompleteness?\\
\Oe\ & $R=10$--$14$\,kpc, $Z\sim4$\,kpc &0.33& in thick disc or halo?\\
\Dc\ & $R=9$--$13$\,kpc, $Z\sim-0.3$\,kpc &-0.25& {unclear}\\
\hline\hline
\end{tabular}
\begin{enumerate}
	\item[*] Maximum relative residuals of the substructures.\\
\end{enumerate}
\end{table*}

\begin{figure}
\centering
\includegraphics[scale=0.4]{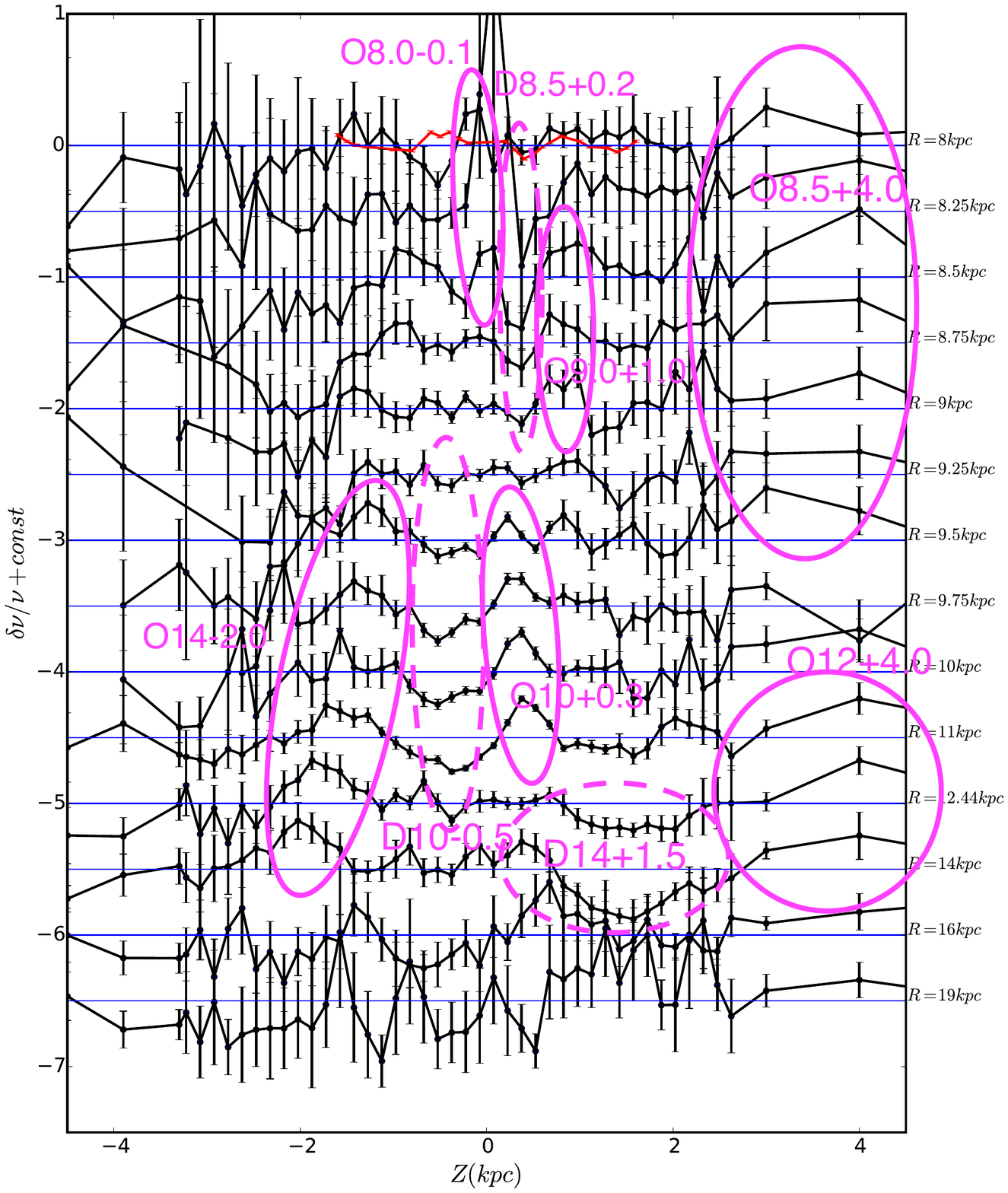}
\vspace{-1cm}
\caption{The black lines show residuals of the vertical stellar density profiles, $\delta\nu/\nu\equiv(\nu(Z)-\nu_{model}(Z))/\nu_{model}(Z)$, as functions of $Z$. The blue lines indicate the zero points corresponding to the residuals at various $R$ bins. The red line indicates the residual profile from~\citet{widrow2012}. The magenta solid and dashed ellipses indicate the approximated locations of the overdensities and dips, respectively.}\label{fig:residual2}	
\end{figure}

\begin{figure}
	\centering
 	\includegraphics[scale=0.3]{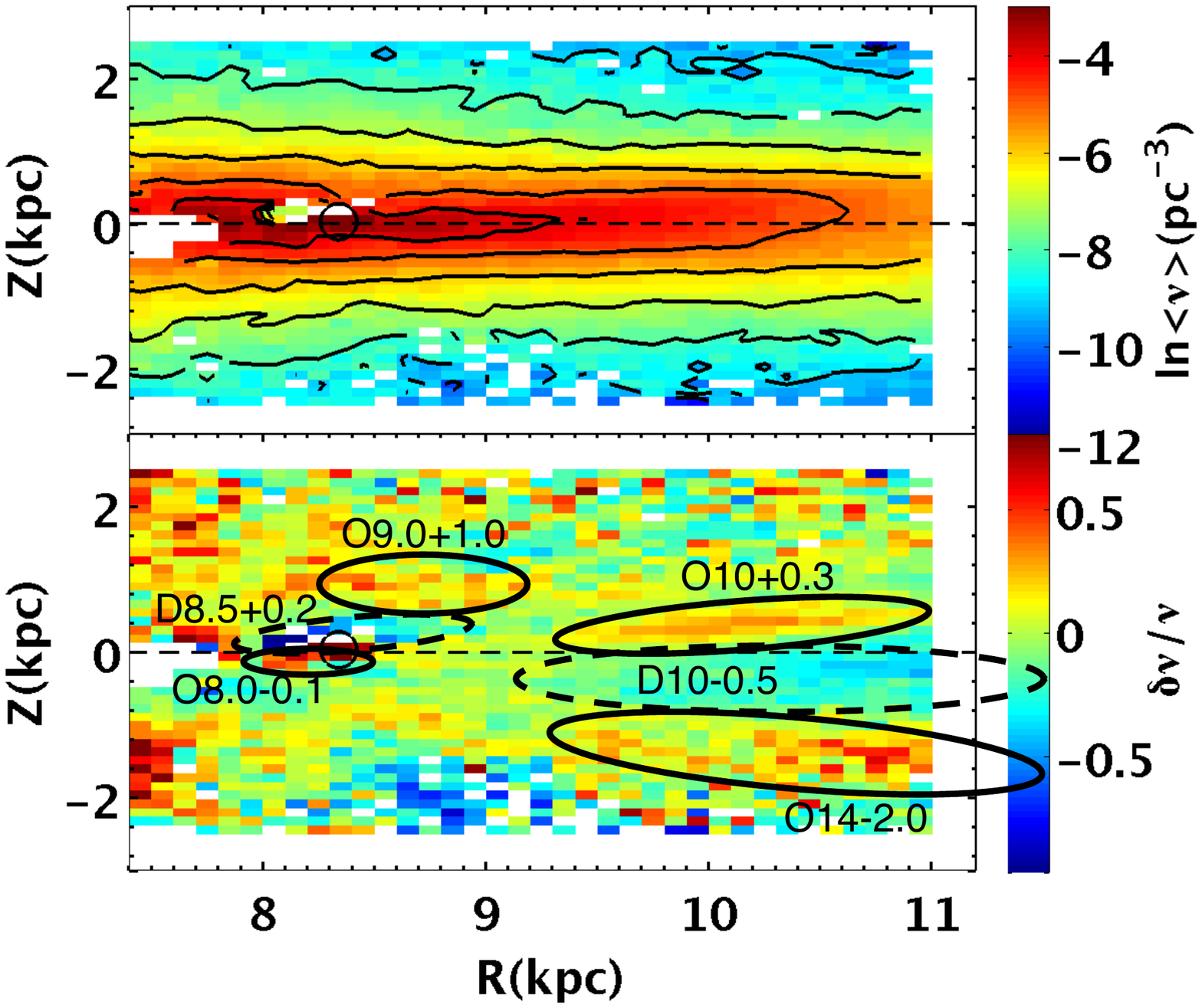}
 	\vspace{-2cm}
 	\caption{The figure is similar to Figure~\ref{fig:RZdensityresidual} but zoomed in to $7<R<11$\,kpc.}
    \label{fig:RZdensityresidual_zoomin}
\end{figure}

Figure~\ref{fig:residual2} shows the residuals of the vertical density model at various $R$ bins. It is obviously seen that the residuals do not randomly fluctuate but show lots of substructures, including overdensities and dips. 

To emphasise these substructures in $R$--$Z$ plane, we produce a 2D density model by interpolating the 1D vertical density models along $R$. Then the 2D residual map (bottom panel of Figure~\ref{fig:RZdensityresidual}) can be derived by subtracting the model from the data, i.e. $\delta\nu/\nu\equiv(\langle\nu\rangle(R,Z)-\nu_{model}(R,Z))/\nu_{model}(R,Z)$. The bottom panel of Figure~\ref{fig:RZdensityresidual_zoomin} shows the similar residual map but is zoomed in to smaller region around the Sun. {It is noted that there are white holes in the upper panel. They are because the lack of targets in these small regions in the LAMOST survey.}

Together with Figures~\ref{fig:RZdensityresidual},  \ref{fig:residual2} and~\ref{fig:RZdensityresidual_zoomin}, we are able to identify 6 overdensities and 3 dips using the criteria that 1) the 1D residuals in Figure~\ref{fig:residual2} are larger than 0 by at least 2-$\sigma$ and 2) the features are seen in at least two neighbouring $R$ bins. We nominate the overdensities starting with letter ``O'' and followed with the typical values of $R$ and $Z$. For the dips, the name is started with ``D''. Table~\ref{tab:substructure} lists all 9 identified substructures. We also mark them in Figures~\ref{fig:RZdensityresidual}, \ref{fig:residual2} and~\ref{fig:RZdensityresidual_zoomin} using solid ellipses for overdensities and dashed ellipses for dips. We discuss them individually in next subsections.

\subsubsection{Overdensity \Ob}
The overdensity \Ob\ is clearly shown in Figures~\ref{fig:residual2} and~\ref{fig:RZdensityresidual_zoomin}. Since the stars contributed to this overdensity are not located in low Galactic latitude, the extinction should not affect the density measurement. The location is consistent with the overdensity found at $R\sim9.5$\,kpc and $Z\sim0.8$\,kpc by \citet{juric2008} (see the second and third rows in the right-most column of their Figure~26 and the right panel of their Figure~27). 

\subsubsection{Overdensity \Od}
The overdensity \Od\ is prominent in Figures~\ref{fig:residual2} and~\ref{fig:RZdensityresidual_zoomin}. The Galactic latitude for the densest point of \Od, which is located around $10.3$\,kpc, is $\sim15^\circ$. The distance and the latitude for \Od\ is consistent with the north near structure discovered by \xu.

\subsubsection{Overdensity \Oc}
The overdensity \Oc, as seen in Figure~\ref{fig:RZdensityresidual}, is one of the most prominent substructures. The Galactic latitude for the densest point of \Oc, which is located at $R\sim13$\,kpc, is around $-20^\circ$. Considering this angle and that the distance to the Sun is around 5\,kpc, we can identify that it should be the south middle substructure discovered in~\xu. 
Unlike \xu, who identify it from colour-magnitude diagram, the 2D density map shows the spatial details that \Oc\ is substantially elongated by about 6\,kpc in $R$. Such a large range can not be explained by the uncertainty of the distance estimate, which is 20\%. At distance of 5\,kpc, such uncertainty can only produce an elongation feature with length of $2$\,kpc.

\subsubsection{Dip \Dd}
The dip \Dd\ is substantially seen in Figures~\ref{fig:RZdensityresidual} and \ref{fig:residual2}. Because the stars contributed to this feature are mostly located in the middle Galactic latitude, the effect of the interstellar extinction can be negligible. Therefore, it is likely a real dip in the outer disc. 

\subsubsection{Dip \Da}
The dip \Da\ is consistent with the north dip shown in \citet{widrow2012} at $R\sim8$\,kpc. In Figure~\ref{fig:RZdensityresidual_zoomin}, we find it almost extends to 9\,kpc in the north of the Galactic mid-plane.

Further compared to \cite{widrow2012} at $Z<0$ (the red line shown in Figure~\ref{fig:residual2}), we find that the residuals in our result is not consistent with these authors. It is noted that they measured the vertical density profile using the data located within $54^\circ<|b|<68^\circ$. Consequently, the density at different $Z$ corresponds to slightly different $R$. Because in this work the vertical density is measured in $R$ slices, in which the density at different $Z$ is independent of $R$, the substructures derived by these authors may not be completely same as our result.   
 
\subsubsection{Overdensity \Oba}
The overdensity \Oba\ is located at $Z\sim4$\,kpc, which is too high to be a substructure of the thin disc. Compared to \citet{juric2008}, \Oba\ is just at the right edge of the Virgo overdensity. Hence, it may be the outer part of Virgo overdensity.

\subsubsection{Overdensity \Oa}
The overdensity \Oa\ is prominent at $R\sim8.25$\,kpc, which is very close to the Sun. At such small distance to the Sun, the RGB stars may suffers from the incompleteness since their apparent magnitude is close to the brightest limiting magnitude of the LAMOST survey ($r\sim9$\,mag). Hence, it is not clear whether the feature is real. 

\subsubsection{Overdensity \Oe}
The overdensity \Oe\ is also located at $Z\sim4$\,kpc, similar to \Oba\ but at larger distance. Its location does not match the Virgo overdensity. It may be either a new substructure in the thick disc/halo or a noise due to the very few samples at larger $Z$.

\subsubsection{Dip \Dc}
{Because the dip \Dc\ is very close to the Galactic mid-plane, it may be more affected by the interstellar extinction than other substructures.}
Therefore, it is not clear {the nature of} this substructure.

\subsubsection{Monoceros ring}\label{sect:mon}
\begin{figure}
\centering
\includegraphics	[scale=0.5]{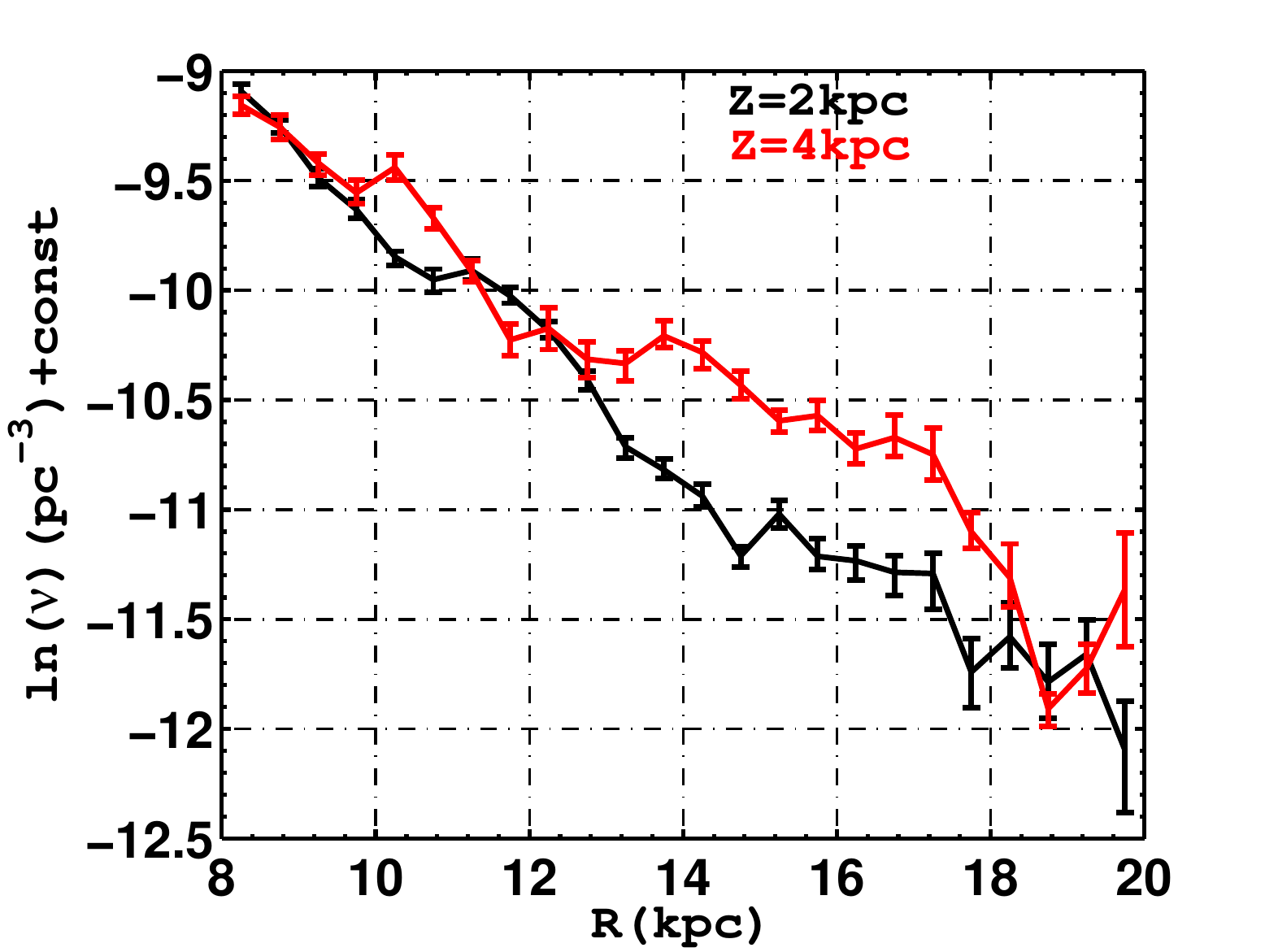}
\caption{The black and red data points stand for the stellar density profiles at $Z=+2$ and $+4$\,kpc, respectively.}\label{fig:checkmon}
\end{figure}
Because that the RGB samples can complete as far as $\sim20$\,kpc, they should be able to exhibit the Monoceros ring located at $8$--$10$\,kpc away from the Sun. However, in the bottom panels of Figures~\ref{fig:bestfitparams} and~\ref{fig:residual2}, no statistically substantial substructure is identified in the north of the Galactic mid-plane at around $R\sim16$--$19$\,kpc, although the stellar density at $Z\sim+4$\,kpc is slightly larger than that at $\sim-4$\,kpc by $\sim1$-$\sigma$.

{The absence of the Monoceros ring is very likely because that this substructure is just a part of the flare, which has been taken into account in our disc models. In Figure~\ref{fig:checkmon}, we demonstrate how the Monoceros ring shows itself up when the flare is neglected. It shows that the radial density profile at $Z=4$\,kpc (red line) exhibits a bump at $R>14$\,kpc compared to the radial density profile at $Z=2$\,kpc (black line). When flare is not taken into account, the disc follows a perfect radial exponential profile, which implies that the scale length of the radial profile measured at any $Z\neq0$ should be same as that at $Z=0$. Then the bump displayed in the profile at $Z=4$\,kpc would be mistakenly treated as the substructure. When flare is considered in the disc model, the density profile at different heights should show different slope, the larger the $Z$, the flatter the radial profile. This can naturally explain the bump at $R>14$ showing in Figure~\ref{fig:checkmon}.}

{This inference is essentially consistent with \citet{momany2006} and some other works. Note that both \citet{juric2008} and \xu\ did not take into account the effect of the flare in the outskirt. Thus, it is not surprising that they can see the overdensity at larger $Z$ by comparing with the exponential model.}


In general, a flare should show north-south symmetry. However, if the mid-plane of the (thick) disc slightly shift to $Z>0$, the stellar density in north may be larger than that in south, as discussed in section~\ref{sec:asymmetry}. The moderate asymmetry shown in the residual of the vertical density profile at $R=19$\,kpc may imply that the mid-plane of the disc shifts to north to some extent. This could be the reason that previous works found that the Monoceros ring is more prominent in the north than in the south~\citep[][and \xu]{newberg2002}. However, from the current RGB stars, although the north--south asymmetry can be barely seen, it is not statistically significant. Moreover, the larger interstellar extinction in the south of the Galactic mid-plane may induce completeness issue in the larger distance. This may result in a systematic underestimation of the stellar density in the south side of the Galactic mid-plane.


\section{Discussions}\label{sec:disc}

\subsection{Assess the systematic bias induced by the broad $R$ bins}\label{sec:systematics}
\begin{figure}
\centering\includegraphics[scale=0.5]{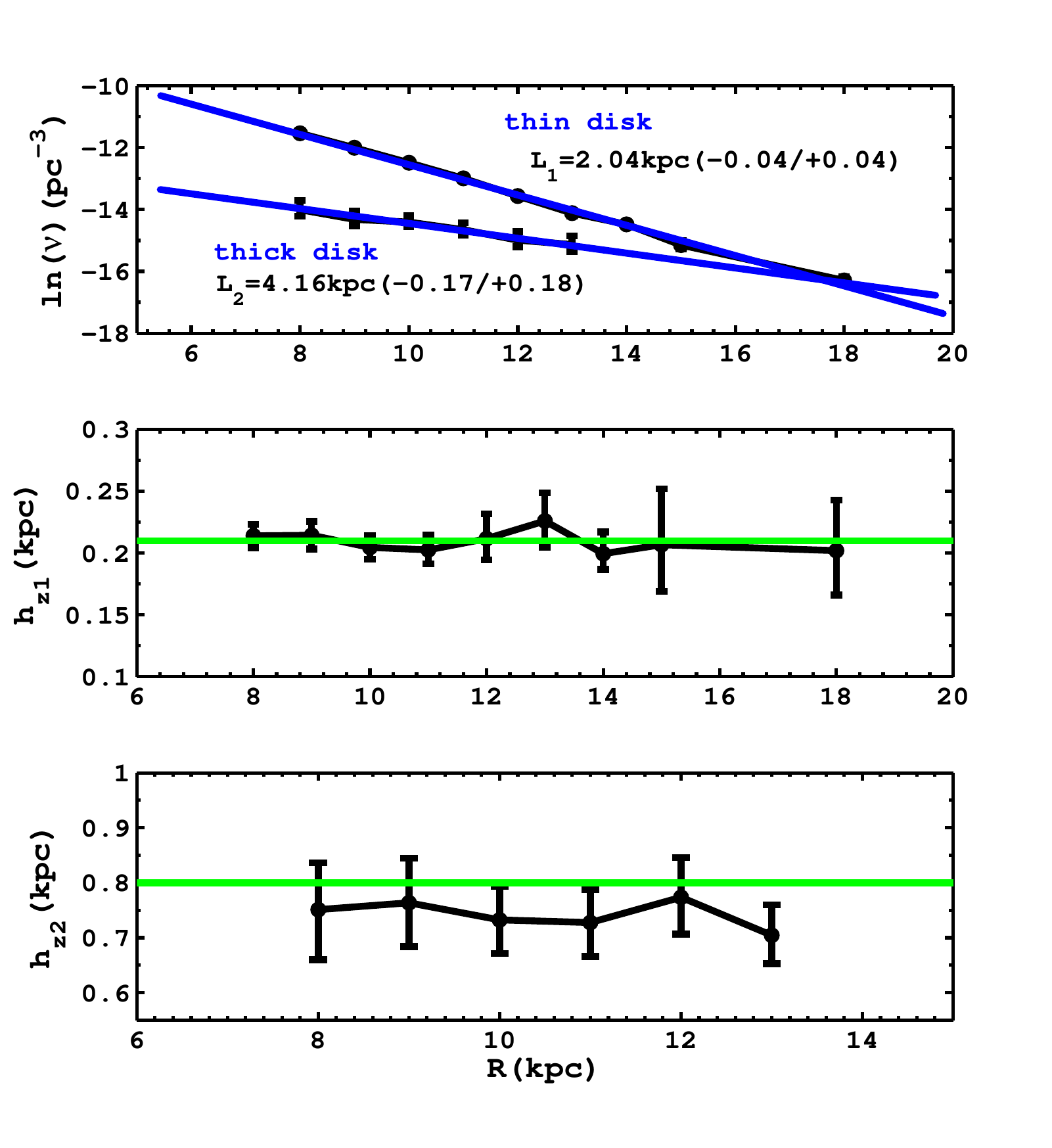}	
\caption{The top panel shows the derived radial density profile using the mock dataset drawn from a toy model of the Galaxy. The radial profile of the thin disc is drawn with black dots and located on the top of the panel, while the radial profile of the thick disc is drawn at the bottom of the panel. The two blue lines are the best-fit exponential models for the two disc components. The middle panel shows the derived scale heights of the thin disc with black dots. The green horizontal line indicates the pre-defined scale height for the thin disc. The bottom panel is similar to the middle one but for the scale height of the thick disc.}\label{fig:validate}
\end{figure}

Ideally, the width of $R$ bins should be infinitely small so that the vertical stellar density in this bin would not be affected by radial difference of the stellar density within the bin. However, in practice, the size of the $R$ bins in this work changes from $0.25$ to $4$\,kpc, which is comparable to the scale lengths of the discs. Therefore, it is worthy to assess whether the binning technique induces any systematic bias in the determination of the structural parameters.  

For this purpose, we apply the modelling method to a mock dataset and compare the derived parameters with the pre-defined values. {We separate the $R$ bins with much larger sizes than the real data, i.e. $1$\,kpc at $R\leq13$\,kpc and $4$\,kpc at $r=18$\,kpc, so that the potential systematics due to the sizes of the bins can be better demonstrated}. The mock stellar density data is then randomly drawn from the pre-defined Galactic model for each $R$ bin. We keep the similar number of the stellar density data as the real observed data at each $R$ bin.
The mock data follow a pre-defined Galactic model with the scale length of $2$ and $4$\,kpc, respectively, for the thin and thick discs. The scale heights for the thin and thick discs are arbitrarily set at $0.21$ and $0.8$\,kpc, respectively, for all radii. To be simplified, flare is not taken into account in the test dataset. 
The fractions of the thick disc and the halo at the location of the Sun are set to $0.08$ and $0.004$, respectively.  

The derived structural parameters for the mock dataset are displayed in Figure~\ref{fig:validate}. The top panel of the figure shows that the reproduced scale lengths of the thin and thick discs are in well agreement with the pre-defined values. The middle panel shows the derived scale heights of the thin disc at different $R$ bins, well consistent with the pre-defined value. For the thick disc, shown in the bottom panel of the figure, the scale height estimates are around 0.75\,kpc, slightly smaller than the pre-defined value of 0.8\,kpc. Although the underestimations are systematic in the sense that it occurs at all $R$ bins, they are not significant since the offset is smaller than the uncertainty of the scale height estimates, which is around 0.1\,kpc. We therefore conclude that the binning technique in $R$ is reliable in reconstructing the structural parameters of the discs.

\subsection{More investigation in the substructures}\label{sec:asymmetry}
\begin{figure*}
	\centering
	\begin{minipage}{18cm}
	\centering
	\includegraphics[scale=0.5]{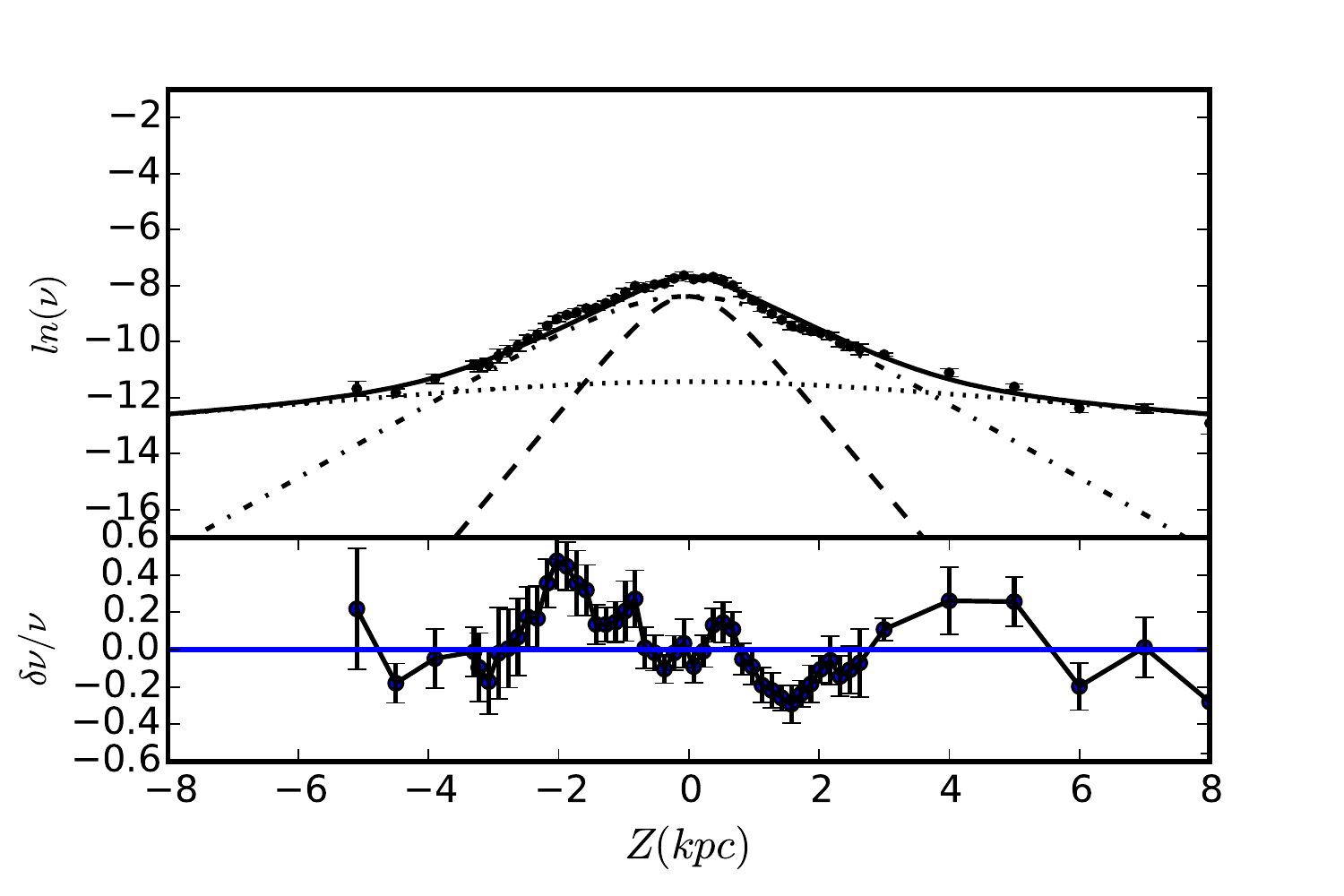}
	\includegraphics[scale=0.5]{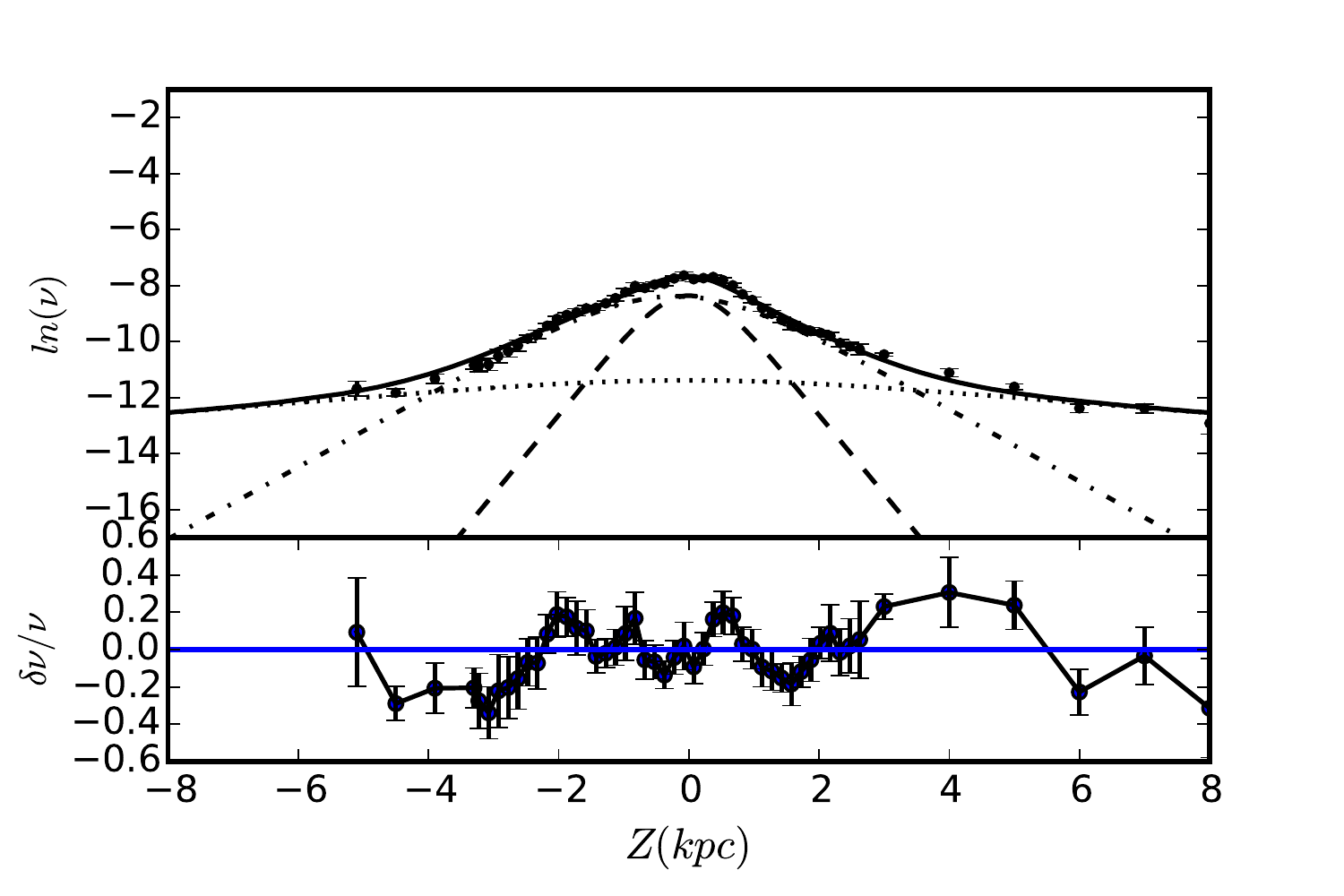}
	\end{minipage}
	\caption{The left panel displays the vertical stellar density profile at $R=14$\,kpc with the best-fit model. The black dots are the observed stellar densities. The dashed, dash-dotted, and dotted lines represent for the thin, thick disc and the stellar halo models, respectively. The solid line stands for the model profile for all components. The bottom shows the relative residual of the model. In the right panel, the dashed and dotted lines represent for the thin disc and the stellar halo models. The dash-dotted line represents for the thick disc with 0.3\,kpc southward shifting. The solid line stands for the model profile for all components, the thin disc, halo and a shifting thick disc. The bottom shows the relative residual of the model with shifting thick disc.}\label{fig:shift14kpc}
\end{figure*}
%

It is noted that, in Figure~\ref{fig:residual2}, residuals at many $R$ bins, especially at $12.44$ and $14$\,kpc, show the dip \Dd\ in the north and the corresponding overdensity \Oc\ in the south. These can be due to the slightly southward offset of the discs. 

\xu\ has considered the effect of the vertically shifting of the thin disc to explain the wave-like substructures in the outer disc. However, the north-south asymmetry stays too high to be explained by the thin disc. We then attempt to explain it by shifting the thick disc, as shown in Figure~\ref{fig:shift14kpc}. The left panel of the figure shows the best-fit model with the thick disc centring at $Z=0$, while the right panel shows the model with a thick disc shifting by $0.3$\,kpc to the south. It is seen that the dip at $Z\sim+1.5$\,kpc and overdensity at $Z=-2$\,kpc displaying in the residual plot of the left panel are significantly weakened in the right panel. Therefore, we suggest that the south-shifting of the thick disc may play a role in producing the substructures \Dd\ and \Oc.%


\subsection{The ripple-like feature in the thick disc at $9<R<10.5$\,kpc}\label{sec:ripple}
\begin{figure}
\centering
\includegraphics[scale=0.5]{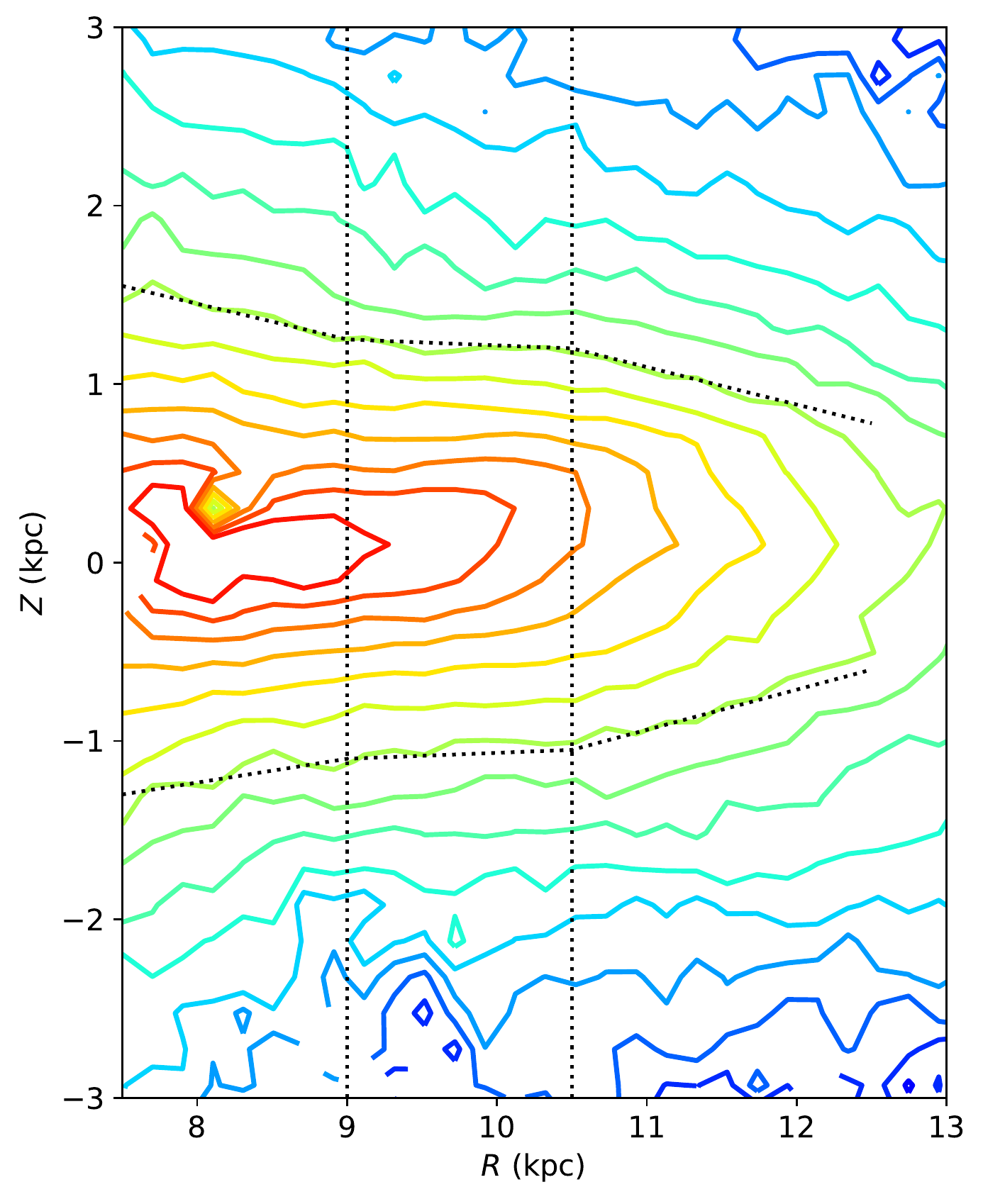}
\caption{The contour shows a smaller region of the mean stellar density map in $R$--$Z$ plane. The dotted broken lines emphasise the trend of the contours at around $|Z|\sim1$\,kpc. The two vertical dotted lines indicate the radii of $R=9$ and $10.5$\,kpc.}\label{fig:9kpc}	
\end{figure}

In Figure~\ref{fig:bestfitparams}, both the radial density and the scale height of the thick disc show ripple-like features at around $R=9$--$10$\,kpc. To test whether this feature is real, we zoom in to the small region at $R=7.5$--$13$\,kpc and $-3<Z<3$\,kpc in the 2D stellar density map and draw the contours in Figure~\ref{fig:9kpc}. At $1<|Z|<2$\,kpc, the $Z$ values of each iso-density contour decline with $R$ at $R<9$ and $R>10.5$\,kpc, while they become flat at $9<R<10.5$\,kpc. Two dotted lines are drawn in both sides of the Galactic mid-plane in the figure to emphasise this broken-contour feature, which may lead to the ripples in the radial stellar density and scale heights of the thick disc.

Two possible reasons may be responsible for the feature. First, the thick disc may be substantially perturbed at this radius. Alternatively, the thick disc is separated into two distinct populations: the inner ($R<9$\,kpc) and the outer ($R>10$\,kpc) thick discs. More investigations is required to clarify the nature of the feature, in particular using the chemical abundance and stellar kinematics.

\section{Summary}\label{sect:summary}
In this work, we use 69\,923 RGB samples carefully selected from the LAMOST DR3 catalogue to reconstruct the spatial structure of the outer disc. The vertical stellar density profiles at various $R$ bins have been fitted with 1D disc-halo models. Then the radial variation of the vertical structural parameters has been displayed without presumption of any analytical function of $R$. We find that
\begin{enumerate}
	\item[(1)]The total radial density profiles for the discs does not follow an exponential profile, but show down-bending and up-bending breaks at $R\sim11$ and $14$\,kpc, respectively. The first down-bending may be due to the drop-down in the radial profile of the thin disc. This is roughly consistent with the prediction of the radial migration. The extrapolation of the radial profile of the thick disc can well overlap with the profile of the outer disc, implying that the outer disc is the extension of the thick disc. This is also supported by the fact that the scale length ($L_{thick}=2.72\pm0.57$\,kpc) of the thick disc is larger than that ($L{thin}=2.13\pm0.23$\,kpc) of the thin disc. Consequently, the thick disc becomes prominent at $R>14$\,kpc, at which the thin disc fades away. This transition may lead to the second up-bending break.\\
	\item[(2)]The flared disc in the outskirt is obviously seen from the increasing scale height at larger radii. Within 12\,kpc, the thin disc moderately increases its scale height, while the thick disc does not show any significant flare in its scale height. The scale heights of the outer disc seems smoothly connected with the scale height of the thick disc at $R=14$\,kpc, hinting that the outer disc may be associated with the thick disc.\\
	\item[(3)]The residual map of the stellar density in $R$--$Z$ plane shows rich substructures, including overdensities and dips. We find that the overdensities \Od\ and \Oc\ likely correspond to the north near and south middle structures, respectively, claimed by \xu. The dip \Da\ is consistent with \citet{widrow2012}. The overdensity \Oba\ may be associated with the Virgo overdensity. The overdensity \Oa\ should be the same substructure located at around $R=9.5$\,kpc found by \citet{juric2008}. Moreover, a dip \Dd, which may correspond to the opposite side of the south middle substructure \Oc, is unveiled from the residual map. Shifting the thick disc model to south may explain the north-south asymmetry produced  by \Dd\ and \Oc.\\
	\item[(4)]We do not find substantial signature for the Monoceros ring in the outer disc. It seems that the substructure can be explained by the flare of the outer disc.\\
	\item[(5)]A strange ripple-like feature located at $9<R<10.5$\,kpc is discovered not only in the density profile and scale heights of the thick disc, but also directly seen in the 2D stellar density map. The nature of this feature is not clear.\\

\end{enumerate}

\section*{Acknowledgements}
We thank Martin L\'opez-Corredoria, Francesca Figueras, and Ivan Minchev for the helpful discussions about this work.
This work is supported by the National Key Basic Research Program of China 2014CB845700. CL acknowledges the NSFC under grants 11373032 and 11333003. Guoshoujing Telescope (the Large Sky Area Multi-Object Fiber Spectroscopic Telescope LAMOST) is a National Major Scientific Project built by the Chinese Academy of Sciences. Funding for the project has been provided by the National Development and Reform Commission. LAMOST is operated and managed by the National Astronomical Observatories, Chinese Academy of Sciences.


\bsp	
\label{lastpage}

\begin{thebibliography}{99}
\bibitem[\protect\citeauthoryear{Astraatmadja \& Bailer-Jones}{2016}]{astraatmadja2016} Astraatmadja T.~L., Bailer-Jones C.~A.~L., 2016, ApJ, 833, 119
\bibitem[\protect\citeauthoryear{Bensby}{2017}]{bensby2017} Bensby T., 2017, IAUS, 321, 3 
\bibitem[\protect\citeauthoryear{Bird, Kazantzidis, \& Weinberg}{2012}]{bird2012} Bird J.~C., Kazantzidis S., Weinberg D.~H., 2012, MNRAS, 420, 913
\bibitem[\protect\citeauthoryear{Bland-Hawthorn \& Gerhard}{2016}]{blandhawthorn2016} Bland-Hawthorn J., Gerhard O., 2016, ARA\&A, 54, 529 
\bibitem[\protect\citeauthoryear{Bovy et al.}{2012}]{bovy2012c} Bovy J., Rix H.-W., Liu C., Hogg D.~W., Beers T.~C., Lee Y.~S., 2012, ApJ, 753, 148
\bibitem[\protect\citeauthoryear{Bournaud, Elmegreen, \& Martig}{2009}]{bournaud2009} Bournaud F., Elmegreen B.~G., Martig M., 2009, ApJ, 707, L1 
\bibitem[\protect\citeauthoryear{Carlin et al.}{2015}]{carlin2015} Carlin J.~L., et al., 2015, AJ, 150, 4 
\bibitem[\protect\citeauthoryear{Carraro et al.}{2010}]{carraro2010} Carraro G., V{\'a}zquez R.~A., Costa E., Perren G., Moitinho A., 2010, ApJ, 718, 683
\bibitem[\protect\citeauthoryear{Carraro}{2015}]{carraro2015} Carraro G., 2015, BAAA, 57, 138
\bibitem[\protect\citeauthoryear{Carraro et al.}{2017}]{carraro2017} Carraro G., Sales Silva J.~V., Moni Bidin C., Vazquez R.~A., 2017, AJ, 153, 99
\bibitem[\protect\citeauthoryear{Chen et al.}{2001}]{chen2001} Chen B., et al., 2001, ApJ, 553, 184 
\bibitem[\protect\citeauthoryear{Conn et al.}{2012}]{conn2012} Conn B.~C., et al., 2012, ApJ, 754, 101
\bibitem[\protect\citeauthoryear{Cui et al.}{2012}]{cui2012} Cui X.-Q., et al., 2012, RAA, 12, 1197 
\bibitem[\protect\citeauthoryear{de Boer, Belokurov, \& Koposov}{2018}]{deboer2018} de Boer T.~J.~L., Belokurov V., Koposov S.~E., 2018, MNRAS, 473, 647 
\bibitem[\protect\citeauthoryear{Deng et al.}{2012}]{deng2012} Deng L.-C., et al., 2012, RAA, 12, 735 
\bibitem[\protect\citeauthoryear{Debattista et al.}{2006}]{debattista2006} Debattista V.~P., Mayer L., Carollo C.~M., Moore B., Wadsley J., Quinn T., 2006, ApJ, 645, 209
\bibitem[\protect\citeauthoryear{D'Onghia et al.}{2016}]{donghia2016} D'Onghia E., Madau P., Vera-Ciro C., Quillen A., Hernquist L., 2016, ApJ, 823, 4 
\bibitem[\protect\citeauthoryear{Feast et al.}{2014}]{feast2014} Feast M.~W., Menzies J.~W., Matsunaga N., Whitelock P.~A., 2014, Natur, 509, 342
\bibitem[\protect\citeauthoryear{Foreman-Mackey et al.}{2013}]{foreman2013} Foreman-Mackey D., Hogg D.~W., Lang D., Goodman J., 2013, PASP, 125, 306
\bibitem[\protect\citeauthoryear{Gaia Collaboration et al.}{2016}]{gaia2016} Gaia Collaboration, et al., 2016, A\&A, 595, A2 
\bibitem[\protect\citeauthoryear{Gilmore \& Reid}{1983}]{gilmore1983} Gilmore G., Reid N., 1983, MNRAS, 202, 1025
\bibitem[\protect\citeauthoryear{G{\'o}mez et al.}{2013}]{gomez2013} G{\'o}mez F.~A., Minchev I., O'Shea B.~W., Beers T.~C., Bullock J.~S., Purcell C.~W., 2013, MNRAS, 429, 159 
\bibitem[\protect\citeauthoryear{G{\'o}mez et al.}{2017}]{gomez2017} G{\'o}mez F.~A., White S.~D.~M., Grand R.~J.~J., Marinacci F., Springel V., Pakmor R., 2017, MNRAS, 465, 3446 
\bibitem[\protect\citeauthoryear{Green et al.}{2014}]{green2014} Green G.~M., et al., 2014, ApJ, 783, 114 
\bibitem[\protect\citeauthoryear{Green et al.}{2015}]{green2015} Green G.~M., et al., 2015, ApJ, 810, 25 
\bibitem[\protect\citeauthoryear{Hammersley \& L{\'o}pez-Corredoira}{2011}]{Lop11} Hammersley P.~L., L{\'o}pez-Corredoira M., 2011, A\&A, 527, A6 
\bibitem[\protect\citeauthoryear{Ho et al.}{2017}]{ho2017} Ho A.~Y.~Q., Rix H.-W., Ness M.~K., Hogg D.~W., Liu C., Ting Y.-S., 2017, ApJ, 841, 40 
\bibitem[\protect\citeauthoryear{Ivezi{\'c} et al.}{2008}]{Ive08} Ivezi{\'c} {\v Z}., et al., 2008, ApJ, 684, 287-325 
\bibitem[\protect\citeauthoryear{Johnston et al.}{2017}]{johnston2017} Johnston K.~V., et al., 2017, arXiv, arXiv:1709.00491
\bibitem[\protect\citeauthoryear{Juri{\'c} et al.}{2008}]{juric2008} Juri{\'c} M., et al., 2008, ApJ, 673, 864-914 
\bibitem[\protect\citeauthoryear{Kazantzidis et al.}{2008}]{kazantzidis2008} Kazantzidis S., Bullock J.~S., Zentner A.~R., Kravtsov A.~V., Moustakas L.~A., 2008, ApJ, 688, 254-276 
\bibitem[\protect\citeauthoryear{Laporte et al.}{2018}]{laporte2018} Laporte C.~F.~P., G{\'o}mez F.~A., Besla G., Johnston K.~V., Garavito-Camargo N., 2018, MNRAS, 473, 1218 
\bibitem[\protect\citeauthoryear{Li et al.}{2012}]{li2012} Li J., Newberg H.~J., Carlin J.~L., Deng L., Newby M., Willett B.~A., Xu Y., Luo Z., 2012, ApJ, 757, 151 
\bibitem[\protect\citeauthoryear{Liu \& van de Ven}{2012}]{liu2012} Liu C., van de Ven G., 2012, MNRAS, 425, 2144
\bibitem[\protect\citeauthoryear{Liu et al.}{2014}]{liu2014} Liu C., et al., 2014, ApJ, 790, 110
\bibitem[\protect\citeauthoryear{Liu et al.}{2017a}]{liu2017a} Liu C., et al., 2017a, RAA, 17, 096 (\paperI)
\bibitem[\protect\citeauthoryear{Liu et al.}{2017b}]{liu2017c} Liu C., Xu Y., Wang H., Wan J., 2017b, arXiv:1712.03977
\bibitem[\protect\citeauthoryear{L\'{o}pez-Corredoira et al.}{2002}]{lopez2002} L\'{o}pez-Corredoira, M., Cabrera-Savers, A., Carz\'{o}n,F., et al. 2002, \ A$\&$A, 394, 883
\bibitem[\protect\citeauthoryear{L{\'o}pez-Corredoira \& Molg{\'o}}{2014}]{lopez2014} L{\'o}pez-Corredoira M., Molg{\'o} J., 2014, A\&A, 567, A106
\bibitem[\protect\citeauthoryear{Luo et al.}{2015}]{luo2015} Luo A.-L., et al., 2015, RAA, 15, 1095 
\bibitem[\protect\citeauthoryear{Majewski, Zasowski, \& Nidever}{2011}]{majewski2011} Majewski S.~R., Zasowski G., Nidever D.~L., 2011, ApJ, 739, 25 
\bibitem[\protect\citeauthoryear{Martig et al.}{2016}]{martig2016} Martig M., Minchev I., Ness M., Fouesneau M., Rix H.-W., 2016, ApJ, 831, 139
\bibitem[\protect\citeauthoryear{Martin et al.}{2004}]{martin2004} Martin N.~F., Ibata R.~A., Bellazzini M., Irwin M.~J., Lewis G.~F., Dehnen W., 2004, MNRAS, 348, 12 
\bibitem[\protect\citeauthoryear{Minchev et al.}{2012}]{minchev2012} Minchev I., Famaey B., Quillen A.~C., Dehnen W., Martig M., Siebert A., 2012, A\&A, 548, A127
\bibitem[\protect\citeauthoryear{Minchev et al.}{2015}]{minchev2015} Minchev I., Martig M., Streich D., Scannapieco C., de Jong R.~S., Steinmetz M., 2015, ApJ, 804, L9 
\bibitem[\protect\citeauthoryear{Minniti et al.}{2011}]{minniti2011} Minniti D., Saito R.~K., Alonso-Garc{\'{\i}}a J., Lucas P.~W., Hempel M., 2011, ApJ, 733, L43
\bibitem[\protect\citeauthoryear{Momany et al.}{2006}]{momany2006} Momany Y., Zaggia S., Gilmore G., Piotto G., Carraro G., Bedin L.~R., de Angeli F., 2006, A\&A, 451, 515 
\bibitem[\protect\citeauthoryear{Newberg et al.}{2002}]{newberg2002} Newberg H.~J., et al., 2002, ApJ, 569, 245 
\bibitem[\protect\citeauthoryear{Pohlen \& Trujillo}{2006}]{pohlen2006} Pohlen M., Trujillo I., 2006, A\&A, 454, 759
\bibitem[\protect\citeauthoryear{Purcell et al.}{2011}]{purcell2011} Purcell C.~W., Bullock J.~S., Tollerud E.~J., Rocha M., Chakrabarti S., 2011, Natur, 477, 301
\bibitem[\protect\citeauthoryear{Quillen \& Garnett}{2001}]{quillen2001} Quillen A.~C., Garnett D.~R.,\ 2001, in Funes J.~G., \& Corsini E. M., eds, ASP,	Conf. Ser. Vol. 230, Galaxy Disks and Disk Galaxies. Astron. Soc. Pac., San Francisco, p. 87
\bibitem[\protect\citeauthoryear{Reyl{\'e} et al.}{2009}]{reyle2009} Reyl{\'e} C., Marshall D.~J., Robin A.~C., Schultheis M., 2009, A\&A, 495, 819 
\bibitem[\protect\citeauthoryear{Rix \& Bovy}{2013}]{rix2013} Rix H.-W., Bovy J., 2013, A\&ARv, 21, 61
\bibitem[\protect\citeauthoryear{Reid et al.}{2014}]{reid2014} Reid M.~J., et al., 2014, ApJ, 783, 130
\bibitem[\protect\citeauthoryear{Ro{\v s}kar et al.}{2008}]{roskar2008} Ro{\v s}kar R., Debattista V.~P., Stinson G.~S., Quinn T.~R., Kaufmann T., Wadsley J., 2008, ApJ, 675, L65 
\bibitem[\protect\citeauthoryear{Skrutskie et al.}{2006}]{skrutskie2006} Skrutskie M.~F., et al., 2006, AJ, 131, 1163
\bibitem[\protect\citeauthoryear{Tian et al.}{2017}]{tian2017} Tian H.-J., et al., 2017, RAA, 17, 114
\bibitem[\protect\citeauthoryear{van der Kruit}{1988}]{vanderkruit1988} van der Kruit P.~C., 1988, A\&A, 192, 117 
\bibitem[\protect\citeauthoryear{van der Kruit \& Freeman}{2011}]{vanderkruit2011} van der Kruit P.~C., Freeman K.~C., 2011, ARA\&A, 49, 301 
\bibitem[\protect\citeauthoryear{Villalobos \& Helmi}{2008}]{villalobos2008} Villalobos {\'A}., Helmi A., 2008, MNRAS, 391, 1806 
\bibitem[\protect\citeauthoryear{Wan et al.}{2015}]{wan2015} Wan J.-C., Liu C., Deng L.-C., Cui W.-Y., Zhang Y., Hou Y.-H., Yang M., Wu Y., 2015, RAA, 15, 1166 
\bibitem[\protect\citeauthoryear{Wan, Liu, \& Deng}{2017}]{wan2017} Wan J.-C., Liu C., Deng L.-C., 2017, RAA, 17, 079 
\bibitem[\protect\citeauthoryear{Wang et al.}{2017}]{wang2017} Wang Q., Wang Y., Liu C., Mao S., Long R.~J., 2017, MNRAS, 470, 2949 
\bibitem[\protect\citeauthoryear{Widrow et al.}{2012}]{widrow2012} Widrow L.~M., Gardner S., Yanny B., Dodelson S., Chen H.-Y., 2012, ApJ, 750, L41  
\bibitem[\protect\citeauthoryear{Xu et al.}{2015}]{xu2015} Xu Y., Newberg H.~J., Carlin J.~L., Liu C., Deng L., Li J., Sch{\"o}nrich R., Yanny B., 2015, ApJ, 801, 105 (\xu)
\bibitem[\protect\citeauthoryear{Xu et al.}{2018}]{xu2017} Xu Y., et al., 2018, MNRAS, 473, 1244 (\paperII) 
\bibitem[\protect\citeauthoryear{Yanny et al.}{2003}]{yanny2003} Yanny B., et al., 2003, ApJ, 588, 824
\bibitem[\protect\citeauthoryear{Yoachim \& Dalcanton}{2006}]{yoachim2006} Yoachim P., Dalcanton J.~J., 2006, AJ, 131, 226
\bibitem[\protect\citeauthoryear{Yu \& Liu}{2018}]{yu2017} Yu J., Liu C., 2018, MNRAS, 475, 1093
\bibitem[\protect\citeauthoryear{Zasowski et al.}{2013}]{zasowski2013} Zasowski G., et al., 2013, AJ, 146, 81 
\bibitem[\protect\citeauthoryear{Zhao et al.}{2012}]{zhao2012} Zhao G., Zhao Y.-H., Chu Y.-Q., Jing Y.-P., Deng L.-C., 2012, RAA, 12, 723 
\end{thebibliography}
\end{document}